\PassOptionsToClass{aps}{revtex4-1}
\documentclass[apj]{emulateapj}
\usepackage{psfig,amsfonts,amsmath,graphicx,natbib,nicefrac,url,hyperref,longtable,enumitem}

\usepackage{booktabs}
\hypersetup{colorlinks=true, urlcolor=blue, citecolor=blue}

\newcommand{\me}{GRB~160509A}

\newcommand{\Fermi}{\textit{Fermi}}
\newcommand{\Swift}{\textit{Swift}}

\newcommand{\EK}{\ensuremath{E_{\rm K}}}
\newcommand{\EKiso}{\ensuremath{E_{\rm K,iso}}}
\newcommand{\Egamma}{\ensuremath{E_{\gamma}}}
\newcommand{\Egammaiso}{\ensuremath{E_{\gamma,\rm iso}}}	     
\newcommand{\epse}{\ensuremath{\epsilon_{\rm e}}}
\newcommand{\epsb}{\ensuremath{\epsilon_{\rm B}}}
\newcommand{\dens}{\ensuremath{n_{0}}}
\newcommand{\Astar}{\ensuremath{A_{*}}}
\newcommand{\tdec}{\ensuremath{t_{\rm dec}}}
\newcommand{\tjet}{\ensuremath{t_{\rm jet}}}
\newcommand{\thetajet}{\ensuremath{\theta_{\rm jet}}}

\newcommand{\AV}{\ensuremath{A_{\rm V}}}

\newcommand{\pcmsq}{\ensuremath{{\rm cm}^{-2}}}
\newcommand{\pcc}{\ensuremath{{\rm cm}^{-3}}}

\newcommand{\nua}{\ensuremath{\nu_{\rm a}}}

\newcommand{\numax}{\ensuremath{\nu_{\rm m}}}

\newcommand{\nuc}{\ensuremath{\nu_{\rm c}}}
\newcommand{\nuaf}{\ensuremath{\nu_{\rm a,FS}}}
\newcommand{\numf}{\ensuremath{\nu_{\rm m,FS}}}
\newcommand{\nucf}{\ensuremath{\nu_{\rm c,FS}}}
\newcommand{\fnumaxf}{\ensuremath{f_{\nu, \rm m,FS}}}
\newcommand{\nuar}{\ensuremath{\nu_{\rm a,RS}}}
\newcommand{\numr}{\ensuremath{\nu_{\rm m,RS}}}
\newcommand{\nucr}{\ensuremath{\nu_{\rm c,RS}}}
\newcommand{\fnumaxr}{\ensuremath{f_{\nu, \rm m,RS}}}

\newcommand{\RB}{\ensuremath{R_{\rm B}}}

\newcommand{\nux}{\ensuremath{\nu_{\rm X}}}

\shorttitle{GRB~160509A}
\shortauthors{}

\def\nrao{1}
\def\ucb{2}
\def\cfa{3}
\def\einsteinfellow{4}
\def\ariz{5}
\def\nyu{6}
\def\ulj{7}
\def\lvjm{8}
\def\bath{9}
\def\ung{10}
\def\mpifr{11}
\def\uw{12}

\submitted{Submitted to ApJ}

\begin{document} 

\title{A Reverse Shock in GRB 160509A}
\shorttitle{A Reverse shock in GRB~160509A}
\shortauthors{Laskar et al.}
\author{
  Tanmoy Laskar\altaffilmark{\nrao,\ucb},
  Kate D.~Alexander\altaffilmark{\cfa},
  Edo Berger\altaffilmark{\cfa},
  Wen-fai Fong\altaffilmark{\einsteinfellow,\ariz},
  Raffaella Margutti\altaffilmark{\nyu},
  Isaac Shivvers\altaffilmark{\ucb},  
  Peter K.~G. Williams\altaffilmark{\cfa},  
  Drejc Kopa{\v c}\altaffilmark{\ulj},  
  Shiho Kobayashi\altaffilmark{\lvjm},
  Carole Mundell\altaffilmark{\bath},    
  Andreja Gomboc\altaffilmark{\ung},  
  WeiKang Zheng\altaffilmark{\ucb},
  Karl M.~Menten\altaffilmark{\mpifr},
  Melissa L.~Graham\altaffilmark{\ucb,\uw},
  and Alexei V. Filippenko\altaffilmark{\ucb}
}
\altaffiltext{\nrao}{National Radio Astronomy Observatory, 520 Edgemont Road, Charlottesville, VA 
22903, USA} 
\altaffiltext{\ucb}{Department of Astronomy, University of California, 501 Campbell Hall, 
Berkeley, CA 94720-3411, USA} 
\altaffiltext{\cfa}{Department of Astronomy, Harvard University, 60 Garden Street, Cambridge, MA 
02138, USA} 
\altaffiltext{\einsteinfellow}{Einstein Fellow}
\altaffiltext{\ariz}{Steward Observatory, University of Arizona, 933 N. Cherry Ave, Tucson, AZ 
85721, USA}
\altaffiltext{\nyu}{Center for Cosmology and Particle Physics, New York University, 4 Washington 
Place, New York, NY 10003, USA}
\altaffiltext{\ulj}{Faculty of Mathematics and Physics, University of Ljubljana, Jadranska 19, 1000 
Ljubljana, Slovenia}
\altaffiltext{\lvjm}{Astrophysics Research Institute, Liverpool John Moores University, 
IC2, Liverpool Science Park, 146 Brownlow Hill, Liverpool L3 5RF, United Kingdom}
\altaffiltext{\bath}{Department of Physics, University of Bath, Claverton Down, Bath, BA2 7AY, 
United Kingdom}
\altaffiltext{\ung}{University of Nova Gorica, Vipavska 13, 5000 Nova Gorica, Slovenia}
\altaffiltext{\mpifr}{Max-Planck-Institut f{\"u}r Radioastronomie, Auf dem Huegel 69, 53121 Bonn, 
Germany}
\altaffiltext{\uw}{Department of Astronomy, University of Washington, Box 351580 U.W., Seattle WA 
98195-1580, USA}


\begin{abstract}
We present the second multi-frequency radio detection of a reverse shock in a $\gamma$-ray burst.
By combining our extensive radio observations of the \Fermi-LAT GRB 160509A at $z = 1.17$ up to 
$20$ days after the burst with \Swift\ X-ray observations and ground-based optical and 
near-infrared data, we show that the afterglow emission comprises distinct reverse shock and forward 
shock contributions: the reverse shock emission dominates in the radio band at $\lesssim10$~days, 
while the forward shock emission dominates in the X-ray, optical, and near-infrared bands. Through 
multi-wavelength modeling, we determine a circumburst density of $\dens\approx10^{-3}$~\pcc, 
supporting our previous suggestion that a low-density circumburst environment is conducive to the 
production of long-lasting reverse shock radiation in the radio band. We infer the presence 
of a large excess X-ray absorption column, $N_{\rm H} \approx 1.5\times10^{22}$~\pcmsq, and a high 
rest-frame optical extinction, $A_{\rm V}\approx3.4$~mag. We identify a jet break in the X-ray light 
curve at $\tjet\approx6$~d, and thus derive a jet opening angle of $\thetajet\approx4\degr$, 
yielding a beaming-corrected kinetic energy and radiated $\gamma$-ray energy of 
$\EK\approx4\times10^{50}$~erg and $\Egamma\approx1.3\times10^{51}$~erg (1--$10^4$~keV, rest frame), 
respectively. Consistency arguments connecting the forward and reverse shocks suggest a deceleration 
time of $\tdec \approx 460$~s~$\approx T_{90}$, a Lorentz factor of $\Gamma(\tdec)\approx330$, and a 
reverse shock to forward shock fractional magnetic energy density ratio of $\RB\equiv\epsilon_{\rm 
B,RS}/\epsilon_{\rm B,FS}\approx8$. 
\end{abstract}

\maketitle

\section{Introduction}
Long duration $\gamma$-ray bursts (GRBs) are produced during the catastrophic collapse of massive 
stars \citep{mw99}, their immense luminosity likely powered by relativistic outflows launched from a 
compact central engine \citep{pir05}. However, the nature of the central engine launching the 
outflow and the mechanism producing the collimated, relativistic jet remain two urgent open 
questions, with models ranging from jets dominated by baryons or by Poynting flux, and those with 
nascent black holes or magnetars providing the central engine \citep[see][for a review]{kz15}.

A direct means of probing the outflow and thus the nature of the central engine is via the study of 
synchrotron radiation from the reverse shock (RS), expected when the ejecta first begin to 
interact with the surrounding medium \citep{mr93,sp99}. Consistency arguments between the 
synchrotron spectrum of the forward shock (FS) and the RS at the time the RS has just crossed the 
ejecta (the deceleration time, $\tdec$) allow a measurement of the ejecta Lorentz factor and the 
ejecta magnetization, i.e., the ratio of the fractional magnetic field energy density of the 
RS-shocked ejecta to that of the FS-shocked circumburst medium.

Theoretically predicted to produce optical flashes on $\sim$ hour timescales, reverse shocks 
were expected to be easily observable with the rapid X-ray localization enabled by \Swift. 
However, this signature has only been seen in a few cases in the \Swift\ era, despite optical 
follow-up observations as early as a few minutes after $\gamma$-ray triggers 
\citep[see][for a review]{jkck+14}.
The dearth of bright optical flashes suggests RS emission may instead be easier to observe at 
longer wavelengths \citep{mmg+07,lbz+13,kmk+15}. We have therefore initiated a program at the 
Karl G. Jansky Very Large Array (VLA) for radio RS studies, and here present the detection of 
a reverse shock in the \Fermi\ \me. 
Combining our radio observations with X-ray data from \Swift\ and ground-based 
optical/near-infrared (NIR) observations, we perform detailed modeling of the 
afterglow in a robust statistical framework to derive the properties of the relativistic ejecta.
Following on GRB~130427A \citep{lbz+13,pcc+14}, this is the second GRB where multi-frequency 
radio observations enable detailed characterization of the RS emission. All magnitudes are
in the AB system \citep{og83}, times are relative to the LAT trigger time, and uncertainties are 
reported at 68\% ($1\sigma$), unless otherwise noted.

\section{GRB Properties and Observations}
\label{text:GRB_Properties_and_Observations}
\subsection{High-energy: \Fermi}
GRB~160509A was discovered by the \Fermi\ Large Area Telescope \citep[LAT;][]{aaa+09} on 2016 
May 09 at 08:59:04.36\,UTC \citep{gcn19403}. 
The burst also triggered the \Fermi\ Gamma-ray Burst Monitor \citep[GBM;][]{gcn19411}. The burst 
duration in the 50--300\,keV GBM band is $T_{90} = 369.7\pm0.8$\,s with a 10\,keV--1\,MeV fluence 
of $(1.790\pm0.002)\times10^{-4}$\,erg\,\pcmsq. 

\subsection{X-ray: \Swift/XRT}
\label{text:data_analysis:XRT}

The \Swift\ X-ray Telescope \citep[XRT;][]{bhn+05} began tiled observations of the LAT error 
circle 2~hr after the GRB. A fading X-ray transient was discovered at RA = 20h\,47m\,00.72s, 
Dec = +76d\,06\arcmin\,28\farcs6 (J2000), with an uncertainty radius of 1\farcs5
\citep[90\% containment;][]{gcn19406,gcn19407, 
gcn19408}.\footnote{\url{http://www.swift.ac.uk/xrt_positions/00020607/}}
The count rate light curve exhibits a break at $\approx4\times10^{4}$\,s. We checked for 
spectral evolution across the break, by extracting XRT PC-mode spectra using the on-line tool on 
the \Swift\ website \citep{ebp+07,ebp+09} 
\footnote{\url{http://www.swift.ac.uk/xrt_spectra/00020607/}} in the intervals $7.3\times10^3$\,s 
to $3.7\times10^4$\,s (spectrum 1) and $4.3\times10^4$\,s to $1.3\times10^6$\,s (spectrum 2). We 
employ \texttt{HEASOFT} (v6.18) and the corresponding calibration files to fit the spectra, assuming 
a photoelectrically absorbed power-law model with the Galactic neutral hydrogen absorption column 
fixed at $N_{\rm H, Gal} = 2.12\times10^{21}~\pcmsq$ \citep{wsb+13}, and tying the value of the 
intrinsic absorption in the host galaxy, $N_{\rm H, int}$, to be the same between the two spectra 
since we do not expect any evolution in the intrinsic absorption with time. We find only marginal 
evidence for spectral evolution, with $\Gamma = 2.01\pm0.05$ in the first spectrum and $\Gamma = 
2.12\pm0.05$ in the second. Fixing the two epochs to have the same spectral index, we obtain 
$\Gamma_{\rm X} = 2.07\pm0.04$ and an intrinsic absorption column, $N_{\rm 
H, int}=(1.52\pm0.13)\times10^{22}$\,\pcmsq. We use this value of $\Gamma_{\rm X}$ (corresponding to 
a spectral index\footnote{We use the convention $f_{\nu}\propto t^{\alpha}\nu^{\beta}$.} of 
$\beta_{\rm X}=-1.07\pm0.04$) and an associated counts-to-flux ratio of 
$6.5\times10^{-11}$\,erg~\pcmsq~s$^{-1}$~ct$^{-1}$ to convert the count-rate to flux density, 
$f_{\nu}$ at 1\,keV.

\subsection{Optical/NIR}
\label{text:data_analysis:optical}
Ground-based observations at Gemini-North beginning at 5.75 hr uncovered a faint source 
($r^{\prime}=23.52\pm0.15$~mag, $z^{\prime}=21.35\pm0.30$~mag) consistent with the XRT 
position \citep{gcn19410}. Subsequent observations by the Discovery Channel Telescope (DCT) $\approx 
1.03$\,d after the LAT trigger showed the source had faded since the Gemini observations, 
confirming it as the afterglow  \citep{gcn19416}. The red color in the Gemini observations, 
$r^{\prime}-z^{\prime}\approx2.1$~mag indicated a high redshift or a significant amount of dust 
extinction within the host galaxy. 

Gemini-North $J$- and $K$-band imaging at $\approx 1.2$~d revealed an NIR counterpart with 
$J\sim16.6$~mag and $K\sim19.7$~mag \citep[Vega magnitudes;][]{gcn19419}.\footnote{In the absence 
of reported uncertainties, we assume an uncertainty of 0.3~mag, corresponding to a $3\sigma$ 
detection.} Spectroscopic observations with Gemini-North at $\approx 1.2$~d yielded a single 
emission line identified as [\ion{O}{2}]3727\AA\ at $z=1.17$, other identifications being ruled out 
by the absence of other lines in the spectrum \citep{gcn19419}. At this redshift, the inferred 
isotropic equivalent $\gamma$-ray energy in the 1--$10^4$ keV rest-frame energy band is 
$\Egammaiso=(5.76\pm0.05)\times10^{53}$\,erg.

We observed \me\ using Keck-I/LRIS \citep{occ+95} beginning at $\approx28.2$~d in $g$- and $R$-band 
with integration times of 972\,s and 900\,s, respectively. We calibrated the data using a custom 
LRIS pipeline, and performed photometry using Starfinder \citep{dbb+00} relative to SDSS stars in 
the field, obtaining $g^{\prime} = 25.39\pm0.12$~mag and $r^{\prime} = 24.18\pm0.35$~mag at 
28.19~d. 

\begin{deluxetable}{ccc}
\tabletypesize{\scriptsize}
\tablecaption{GRB\,160509A: Log of VLA observations \label{tab:data:VLA}}
\tablehead{
\colhead{$\Delta t$} & \colhead{Frequency} & \colhead{Flux density}
\\
\colhead{(d)} & \colhead{(GHz)} & \colhead{($\mu$Jy)}
}
\startdata
0.351 &	  8.5 &	 $43.8 \pm 29.1$ \\
0.351 &	 11.0 &	 $50.6 \pm 27.4$ \\
0.363 &	  5.0 &	 $78.2 \pm 23.9$ \\
0.363 &	  7.4 &	 $90.8 \pm 18.6$ \\
\ldots & \ldots & \ldots
\enddata
\tablecomments{This is a sample of the full table available on-line.}
\end{deluxetable}

\subsection{Radio}
\label{text:data_analysis:radio}
We observed the afterglow with the VLA starting at 0.36\,d. We tracked the flux density 
of the afterglow over multiple epochs spanning 1.2 to 33.5\,GHz, using 3C48, 3C286, and 3C147 
as flux and bandpass calibrators, and J2005+7752 as the gain calibrator. We carried out data 
reduction using the Common Astronomy Software Applications (CASA), and list the results of our VLA 
monitoring campaign in Table \ref{tab:data:VLA}.

\begin{figure*}
\begin{tabular}{cc}
 \centering
 \includegraphics[width=0.45\textwidth]{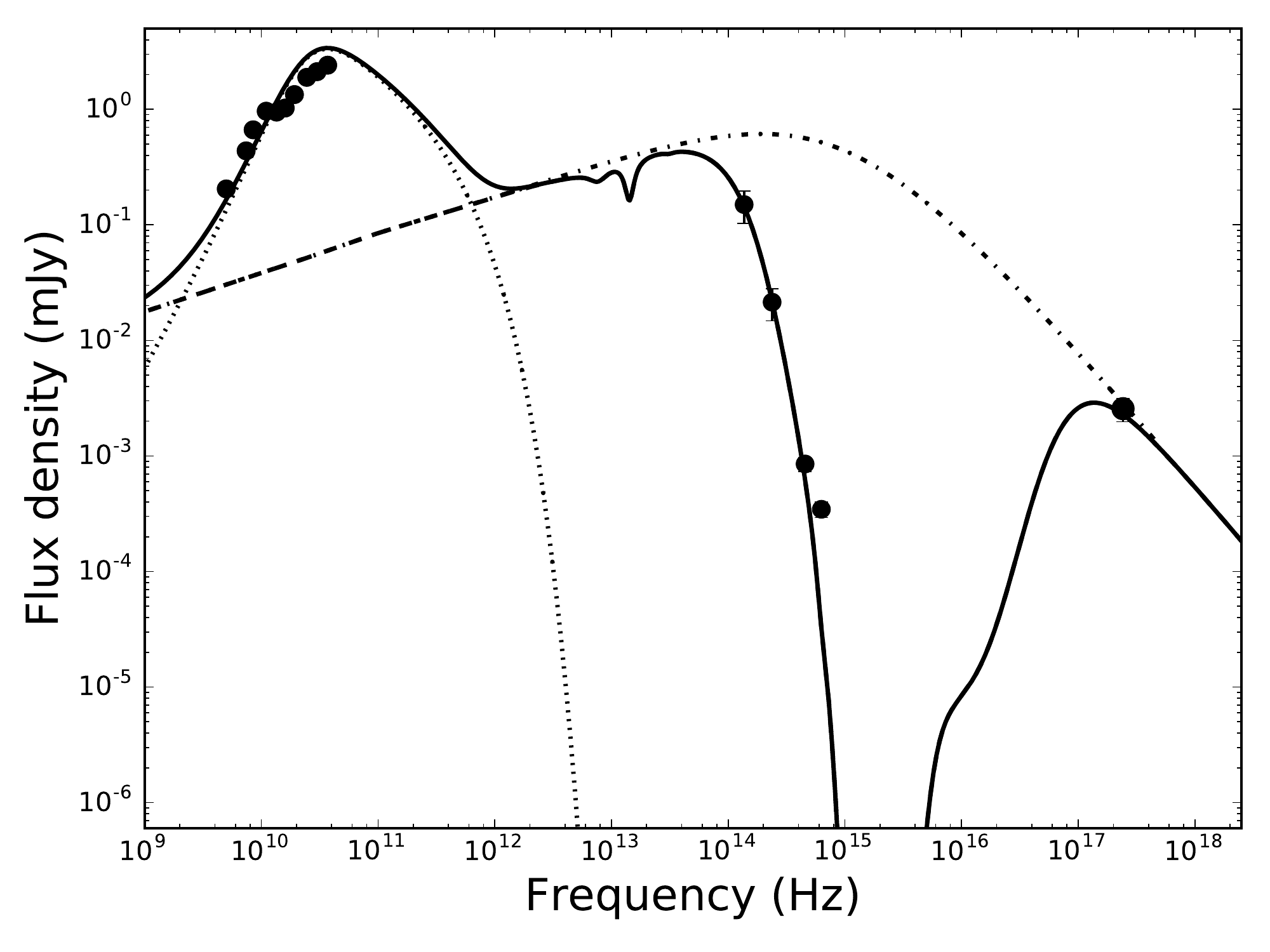} &
 \includegraphics[width=0.45\textwidth]{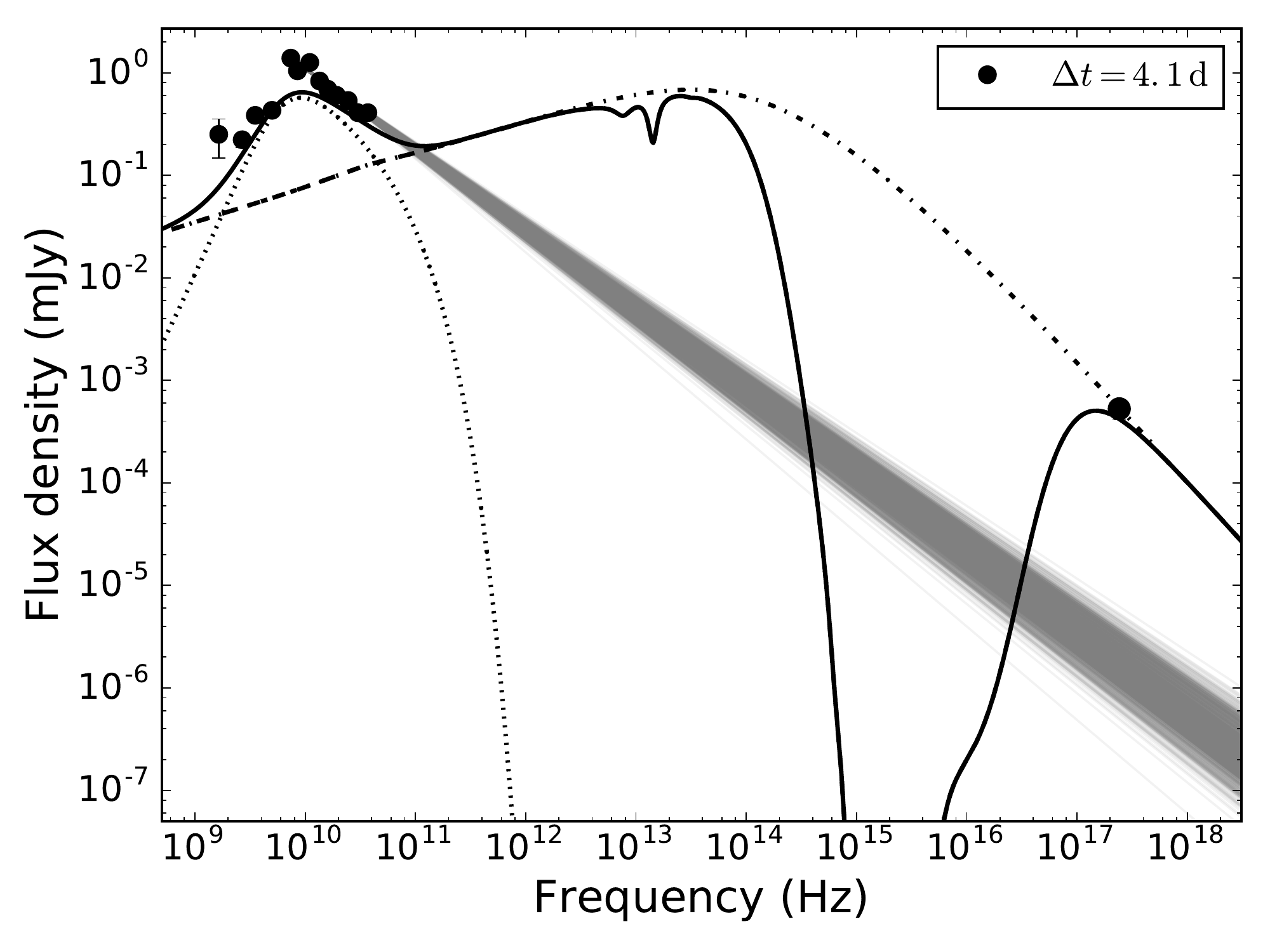} \\
 \includegraphics[width=0.45\textwidth]{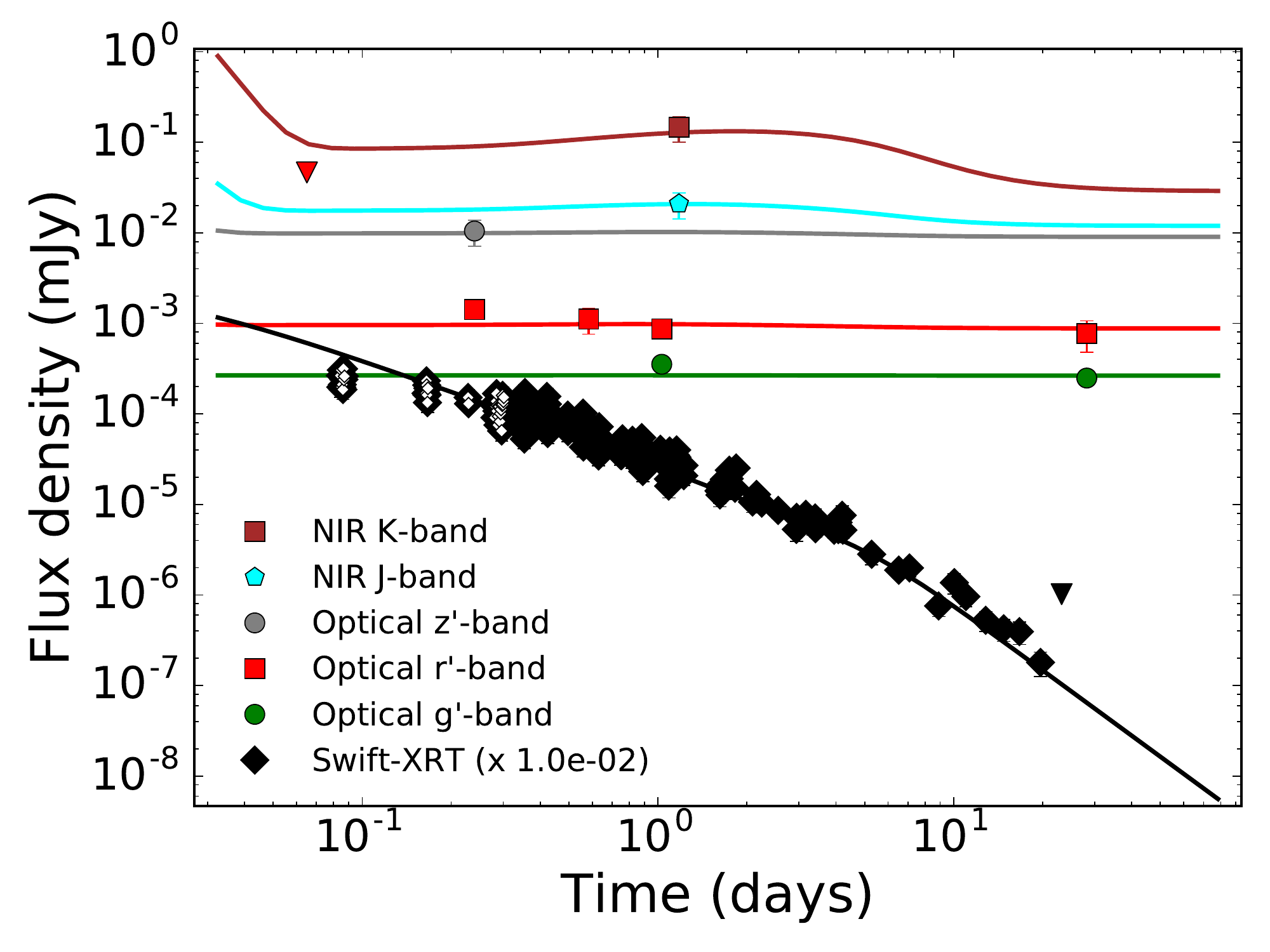} &
 \includegraphics[width=0.45\textwidth]{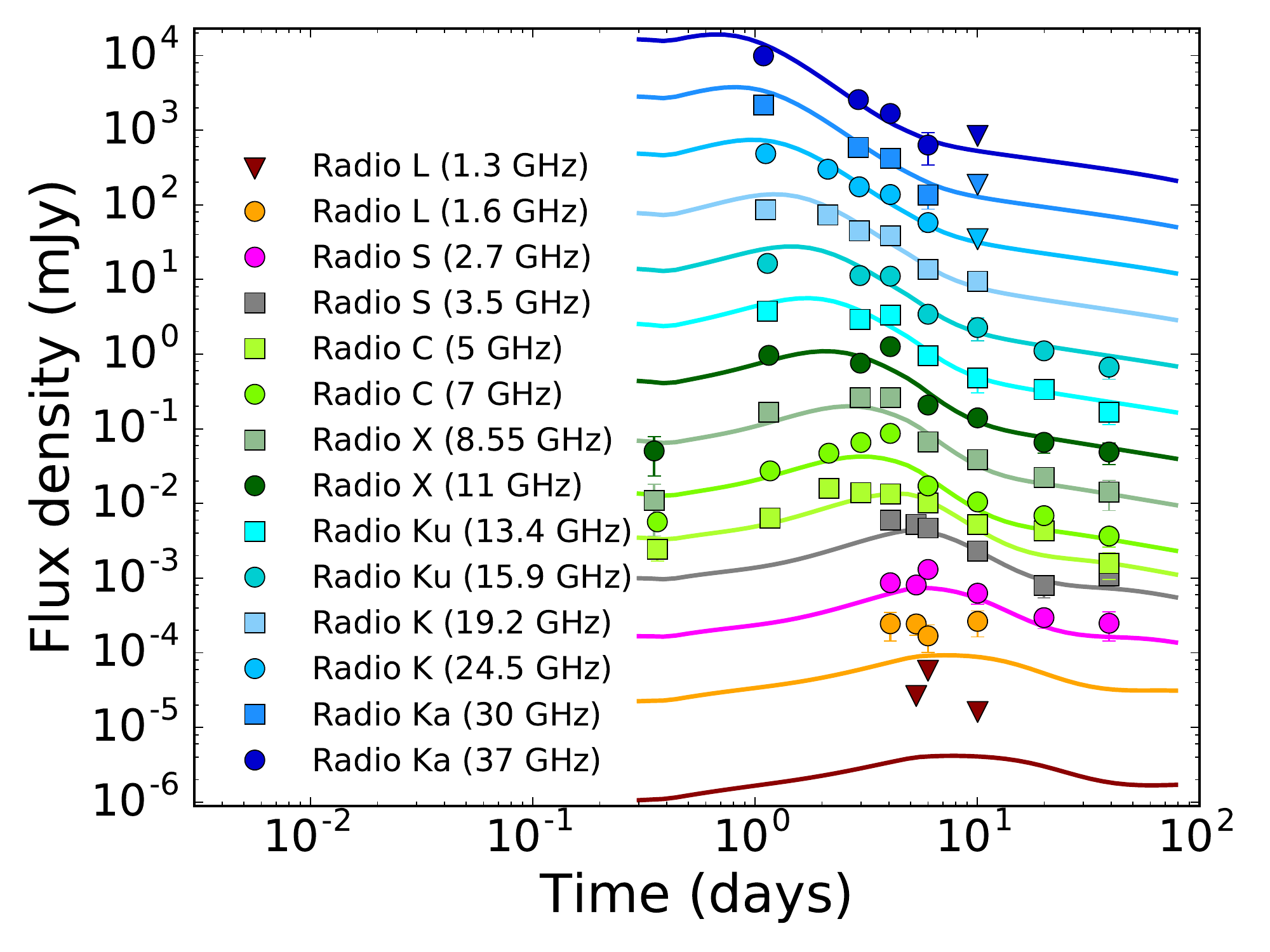}
\end{tabular}
\caption{Top: Radio through X-ray spectral energy distribution of the afterglow of \me\ at 1.1~d 
and 4.1~d (data points), together with a best-fit model (solid lines) comprising 
the forward shock (dashed) and reverse shock (dotted). The dash-dot line indicates the expected 
spectrum of the forward shock in the absence of optical extinction and X-ray photoelectric 
absorption in the host galaxy and in the Milky Way. The optical and NIR points have been 
interpolated to the common time of 1.1~d by a fit to the $r^{\prime}$-band light curve 
($\alpha=-0.33\pm0.02$). The $g^{\prime}$- and $r^{\prime}$-band (and likely also $z$-band) data 
are significantly affected by host flux contribution (Section \ref{text:basic_considerations:opt}). 
The shaded bands in the right panel are a random subset of 1000 MCMC samples from a total of 
$3\times10^{5}$ samples fitting the radio peak with a broken power-law function. The fits to the 
radio data at 4.1~d under-predict the observed X-ray flux at this time by more than two orders of 
magnitude. Bottom: X-ray, optical/NIR (left) and radio (right) light curves using the combined 
best-fit RS+FS model. Adjacent radio light curves have been scaled by factors of 4 for clarity, 
normalized with respect to the light curve at 11~GHz. 
\label{fig:160509A_radioxrtsed}}
\end{figure*}

\section{Multi-wavelength Modeling}
\label{text:modeling}
\subsection{Basic Considerations}
\label{text:basic_considerations}
We interpret the observed behavior of the afterglow from radio to X-rays in the framework 
of the standard synchrotron model, described by three break frequencies (the self-absorption 
frequency, $\nua$, the characteristic synchrotron frequency, $\numax$, and the cooling frequency, 
$\nuc$) and an overall flux normalization, allowing for two possibilities for the density profile of 
the circumburst medium: the ISM profile \citep[$\rho={\rm const}$;][]{spn98} and the wind profile 
\citep[$\rho \propto r^{-2}$;][]{cl00}.

\subsubsection{X-rays -- location of \texorpdfstring{$\nuc$}{the cooling frequency} and a jet 
break}
We fit the \Swift\ XRT light curve as a power-law with two temporal breaks. The first break occurs 
at $t_{\rm b,1} = 0.37\pm0.14$~d when the decline rate steepens from $\alpha_{\rm 
X,1}=-0.51\pm0.12$ to $\alpha_{\rm X, 2} = -1.27\pm0.11$ ($\Delta \alpha_{12} = -0.76\pm0.17$). 
This steepening does not have a simple explanation in the standard synchrotron model 
(for instance, the passage of \nuc\ results in a steepening of the light curve by only $\Delta 
\alpha = -0.25$). It is possible that the X-ray data before $t_{\rm b,1}$ are part of a plateau 
phase, which is commonly observed among GRB X-ray afterglows \citep{nkg+06}, and we therefore do not 
consider the X-ray observations before $\approx0.35$~d in the remainder of our analysis. 

At $t_{\rm b,2} = 5.4\pm2.3$~d, the light curve steepens again to $\alpha_{\rm X,3}=-2.2\pm0.3$ 
($\Delta\alpha_{23}=-0.9\pm0.3$), suggestive of a jet break. Since $\numax \propto t^{-1.5}$ is 
expected to be below the X-ray band at this time and the post-break decay rate at $\nu > \numax$ is 
$t^{-p}$, we determine that the energy index of non-thermal electrons, $p\approx2.2$ \citep{sph99}. 
For this value of $p$, we expect a spectral slope of $\beta_{\rm X} \approx -1.1$ or $\beta_{\rm X} 
\approx -0.6$ for $\nuc < \nux$ and $\nuc > \nux$, respectively. The measured X-ray spectral index 
of $\beta_{\rm X} = -1.07\pm0.04$ requires the former, whereupon we expect $\alpha_{\rm X} = 
(2-3p)/4 \approx -1.2$. This is consistent with the measured value of $\alpha_{\rm X,2} = 
-1.27\pm0.11$. Thus, we conclude that the X-ray light curve and spectrum are both consistent with 
$p\approx2.2$ and $\nuc < \nux$. We note that in this regime the X-ray light curve does not 
distinguish between the ISM and wind models.

\subsubsection{Optical/NIR -- Extinction and Host Flux}
\label{text:basic_considerations:opt}
At the time of the Gemini $z^{\prime}$- and $r^{\prime}$-band observations (0.24~d), the X-ray to 
$z^{\prime}$-band spectral index is flat, $\beta_{\rm ox}=-0.11\pm0.06$, while the 
$z^{\prime}$-$r^{\prime}$ spectral index is extremely steep, $\beta_{\rm zr} = -5.4\pm1.1$. Given 
the moderate redshift of the burst, the only explanation for these observations is a large amount 
of extinction along the sight-line through the GRB host galaxy, suppressing the optical flux. On 
the other hand, the spectral index between the DCT $r^{\prime}$- and $g$-band observations at 
$\approx1$~d is $\beta_{\rm gr}=-1.9\pm0.6$, significantly shallower than $\beta_{\rm zr}$, while 
the $r^{\prime}$-band light curve before $\approx 1$~d declines as $\alpha_{\rm r}=-0.33\pm0.02$,
shallower than expected in the standard afterglow model. Together, these observations indicate a 
significant contribution to the afterglow photometry from the host galaxy. This is confirmed by our 
Keck $g$- and $R$-band observations at $\approx28$~d, which yield flux densities similar to the DCT 
observations at $\approx1$~d. We find that modeling the $r^{\prime}$-band light curve as a sum of a 
power-law and a constant yields $\alpha_{\rm r} = -1.09\pm0.45$, with the additive constant 
$f_{\nu,\rm r} = 0.75\pm0.10$~$\mu$Jy. We note that whereas the light curve decay rate at $\numax < 
\nu < \nuc$ is expected to provide diagnostic power for the circumburst density profile, the paucity 
of optical data and the large uncertainty in the optical decay rate for this event preclude such a 
discrimination. In the detailed modeling (Section \ref{text:FS}) we fit for the host galaxy flux 
density in all optical/NIR filters, together with the optical extinction along the line of sight 
through the host.

\subsubsection{Radio -- Multiple Components}
The radio spectral energy distribution (SED) at 4.06~d exhibits a clear peak at $\approx8.4$~GHz 
with a flux density of $\approx1.2$~mJy. At this time, the measured X-ray flux density is 
$f_{\nu,\rm X} = (6.3\pm1.9)\times10^{-4}$~mJy. Fitting the radio data with a broken power-law and 
extrapolating to the X-rays, we find that the expected X-ray flux density is at least two orders of 
magnitude lower than observed (Figure \ref{fig:160509A_radioxrtsed}).  This suggests that the radio 
and X-ray emission at 4.06~d arise from separate processes. Further, we note that the radio spectral 
index above 10~GHz at 10~d is $\beta_{\rm radio}(10~\rm d)= 0.1\pm0.2$, in contrast to the spectral 
index above the peak at 4.06~d, $\beta_{\rm radio}(4.06~\rm d) = -0.79\pm0.02$. Since such a 
hardening of the spectral index is not expected in the standard synchrotron model, we propose that 
the radio peak at $4.06$~d has faded to reveal a fainter underlying component at 10~d. We show this 
underlying emission to be consistent with the FS in Section \ref{text:FS}.

To summarize, the X-ray spectral index and light curve are consistent with a forward shock 
origin for the X-ray emission with $p\approx2.2$ and $\nuc < \nux$. The radio spectrum at 4.06~d 
cannot be extrapolated to match the observed X-ray flux at this time, suggesting that the radio and 
X-ray emission arise from separate processes. The radio peak at 4.06~d fades to reveal an 
underlying power-law continuum, which we ascribe to the FS. Finally, there is insufficient 
information in the afterglow observations to constrain the circumburst density profile.

\begin{figure*}
\begin{tabular}{ccc}
 \centering
 \includegraphics[width=0.31\textwidth]{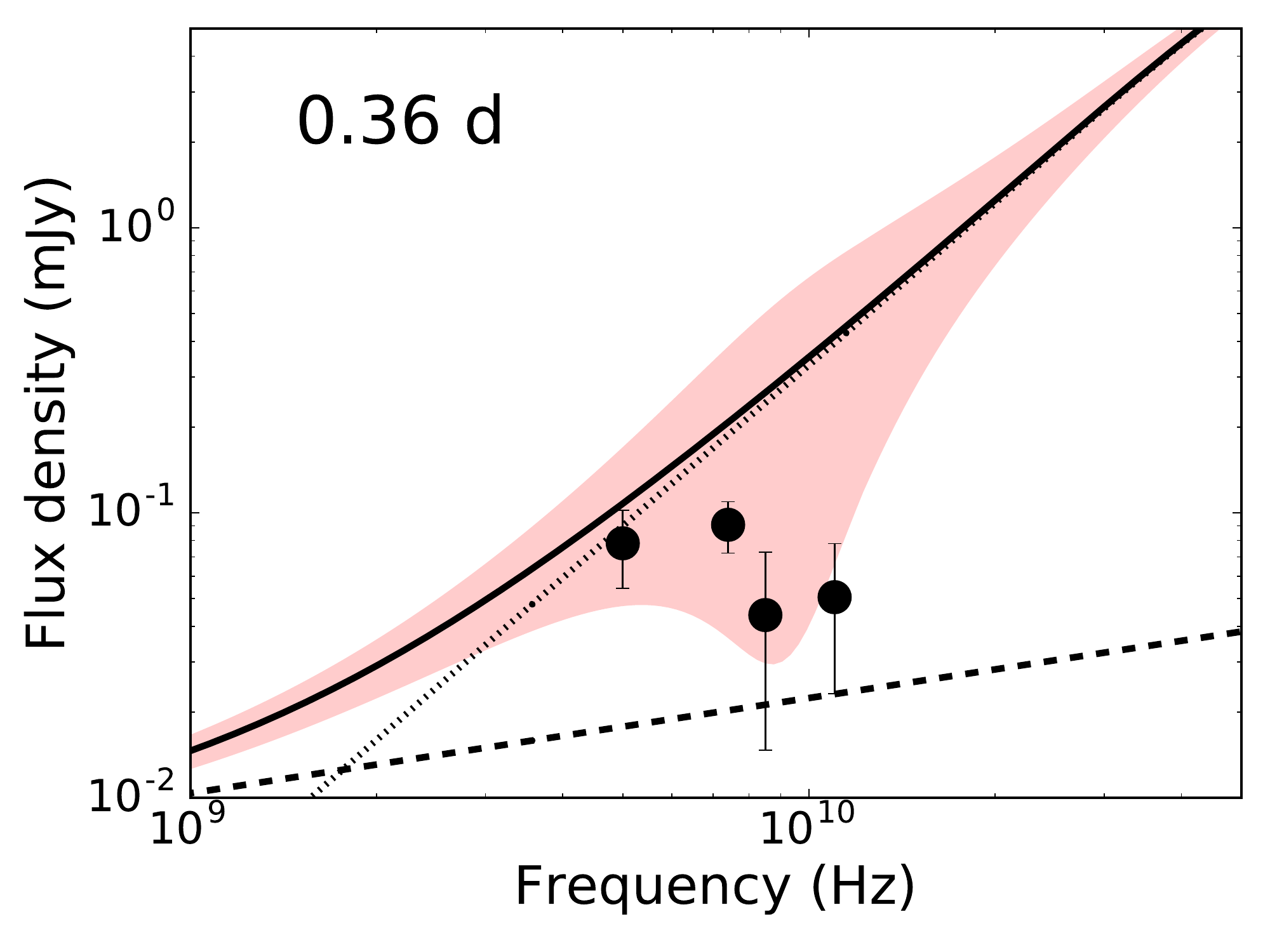} &
 \includegraphics[width=0.31\textwidth]{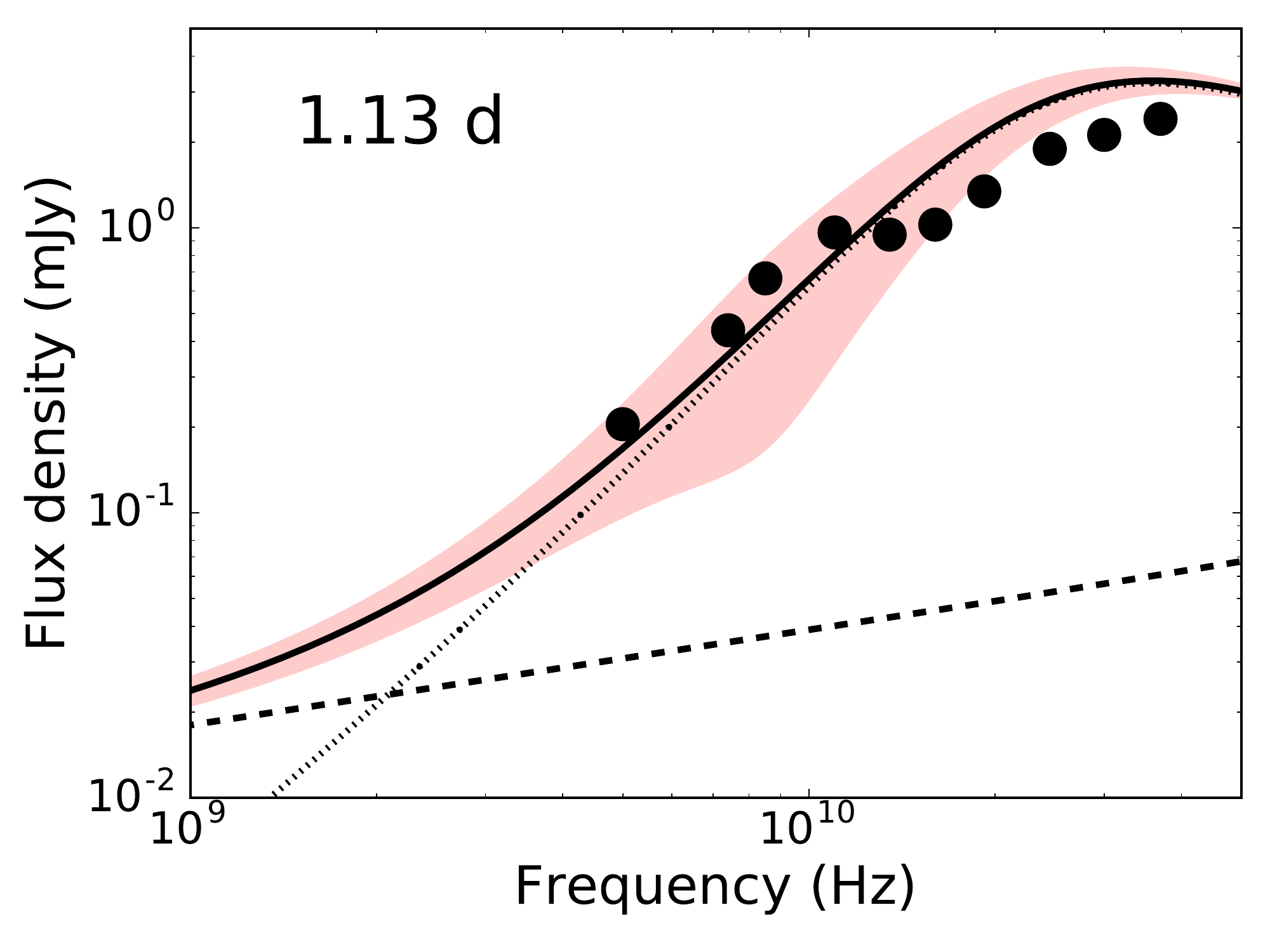} &
 \includegraphics[width=0.31\textwidth]{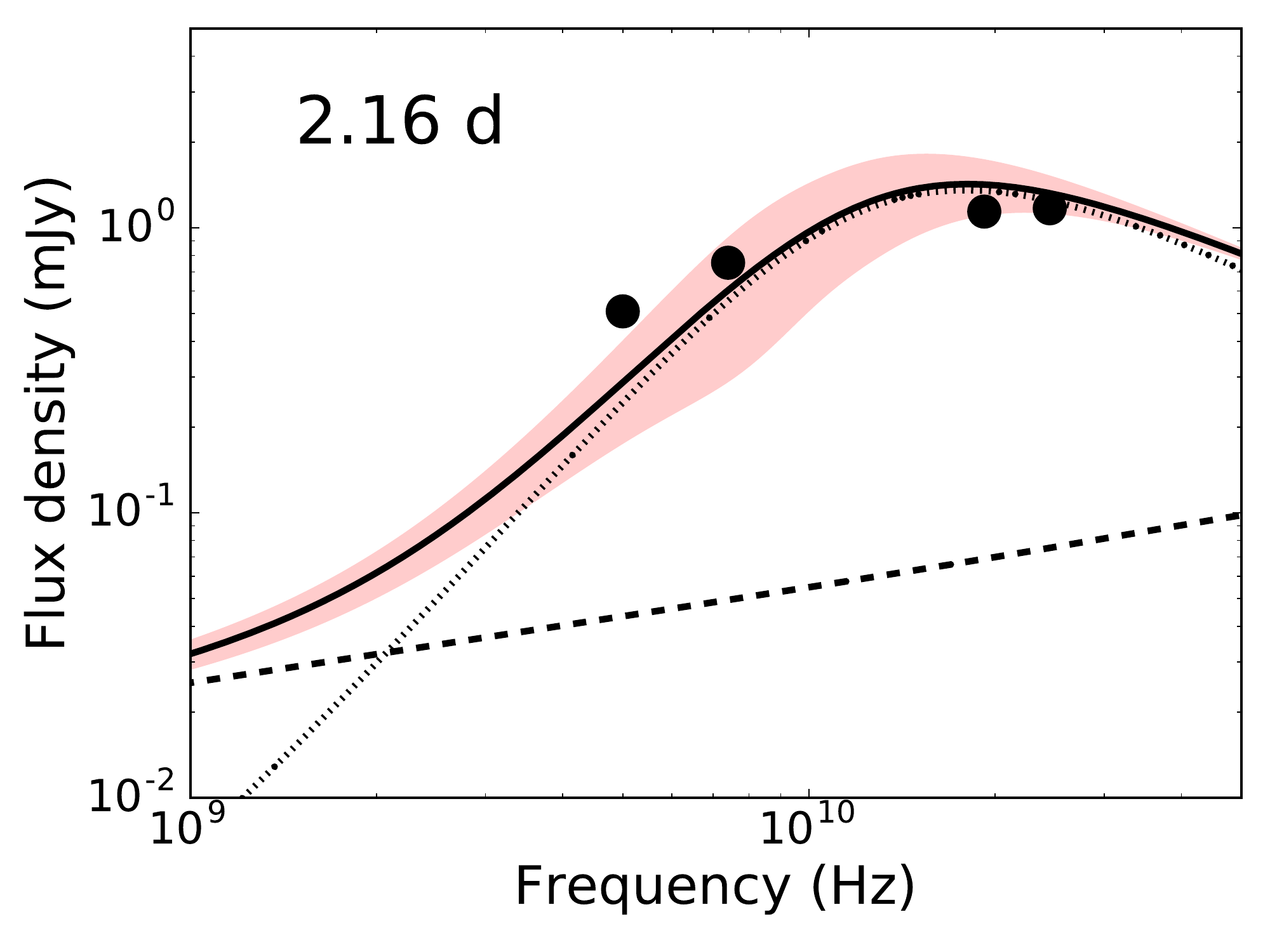} \\
 \includegraphics[width=0.31\textwidth]{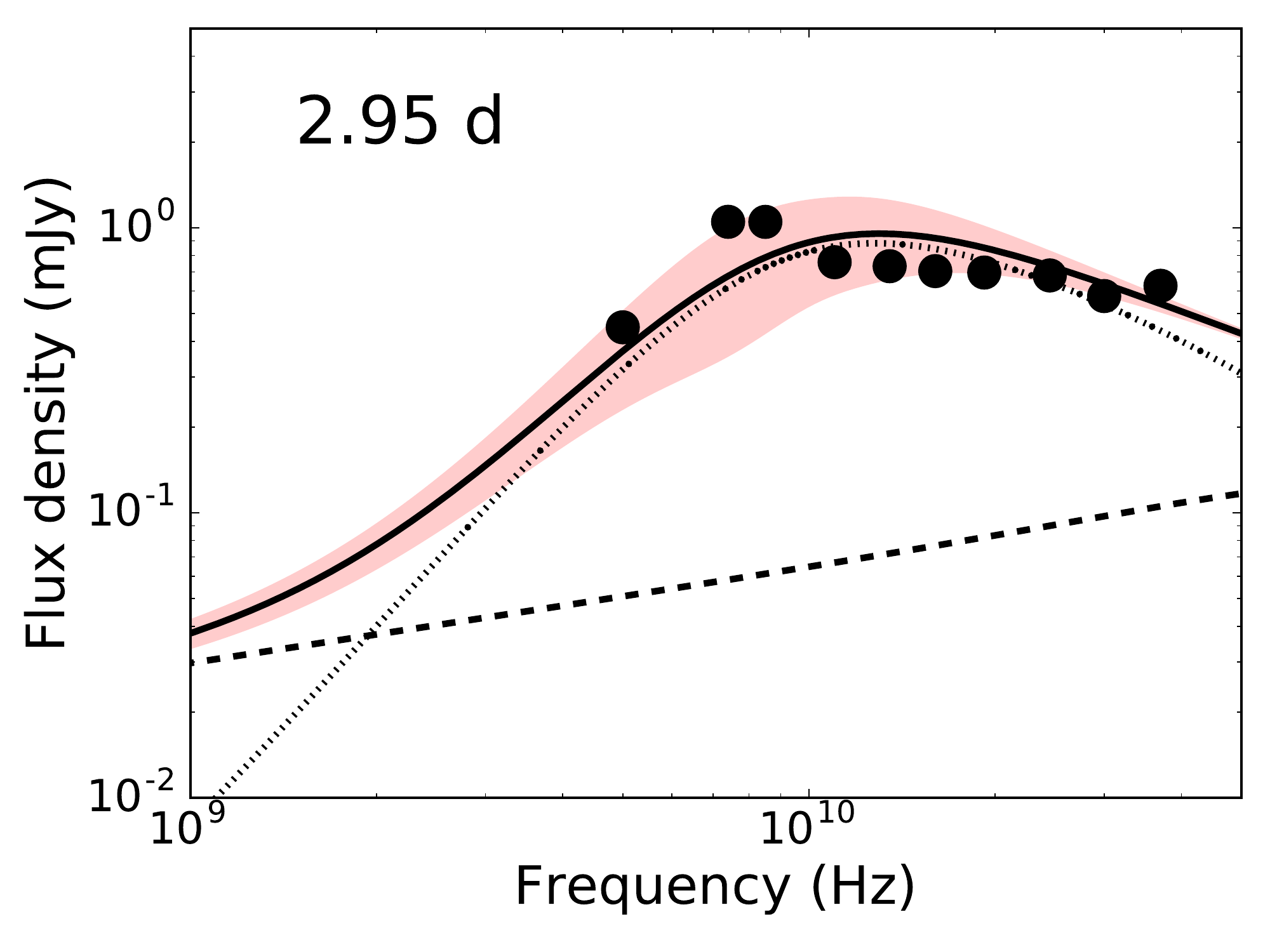} &
 \includegraphics[width=0.31\textwidth]{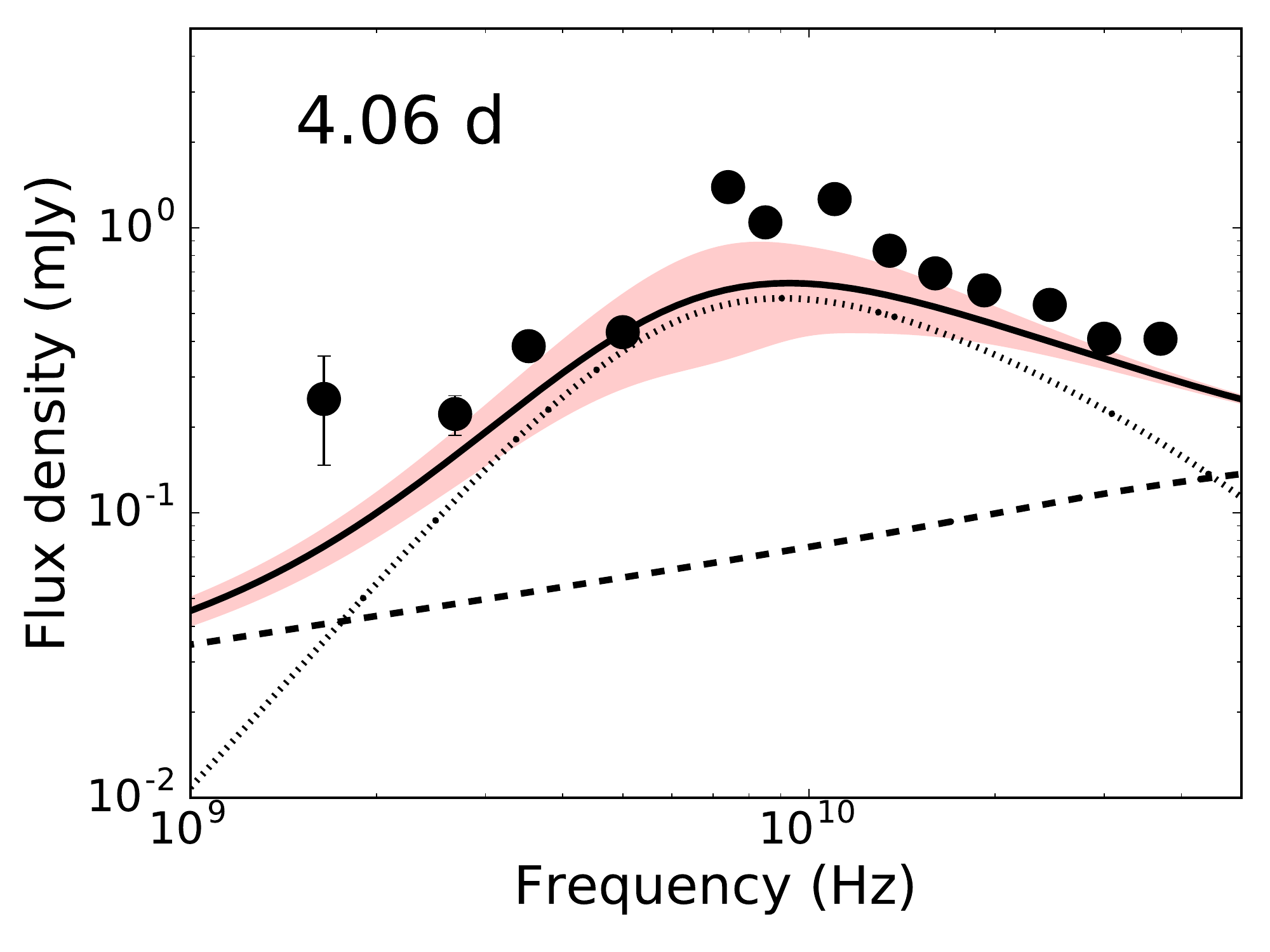} &
 \includegraphics[width=0.31\textwidth]{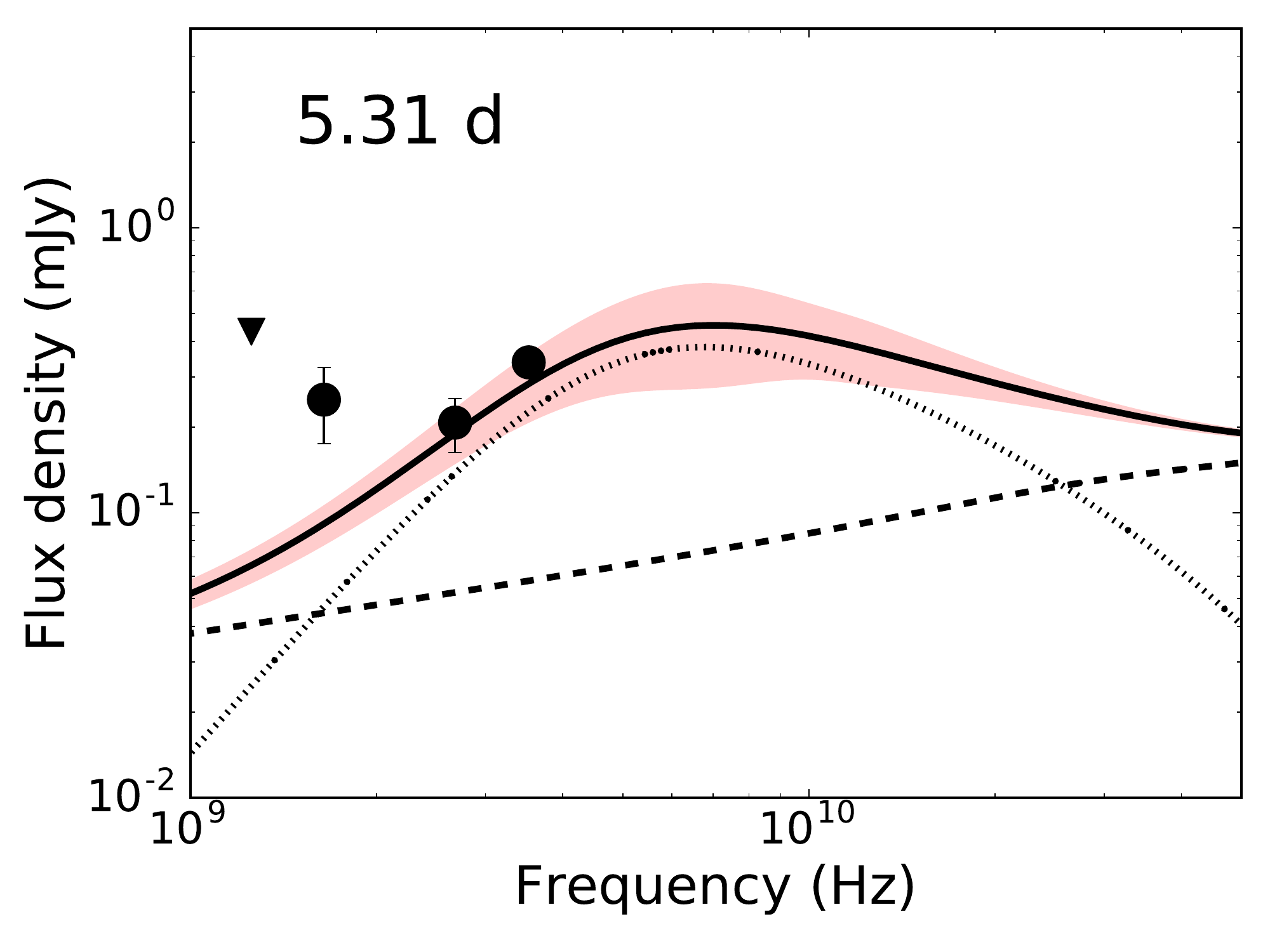} \\
 \includegraphics[width=0.31\textwidth]{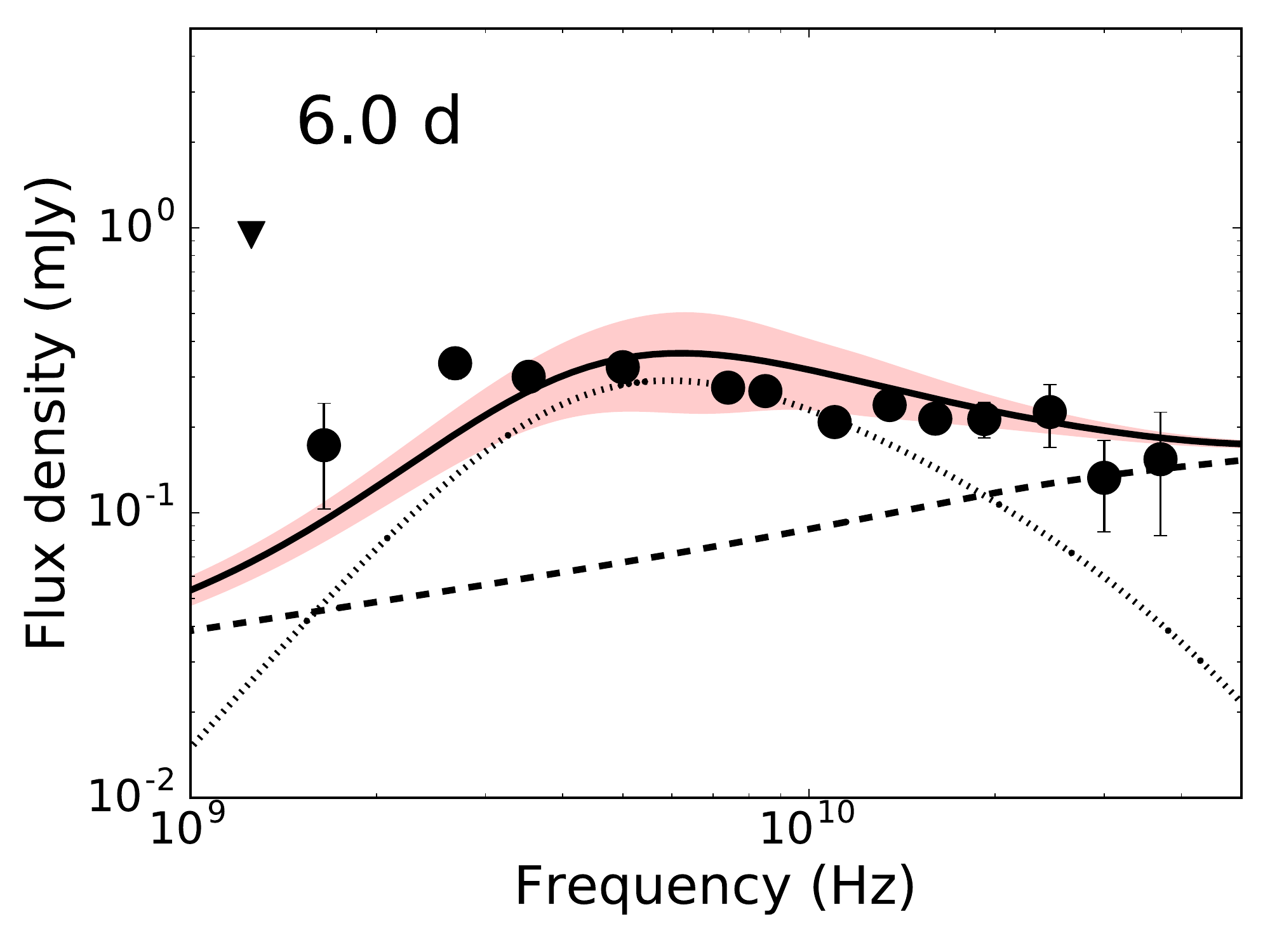} &
 \includegraphics[width=0.31\textwidth]{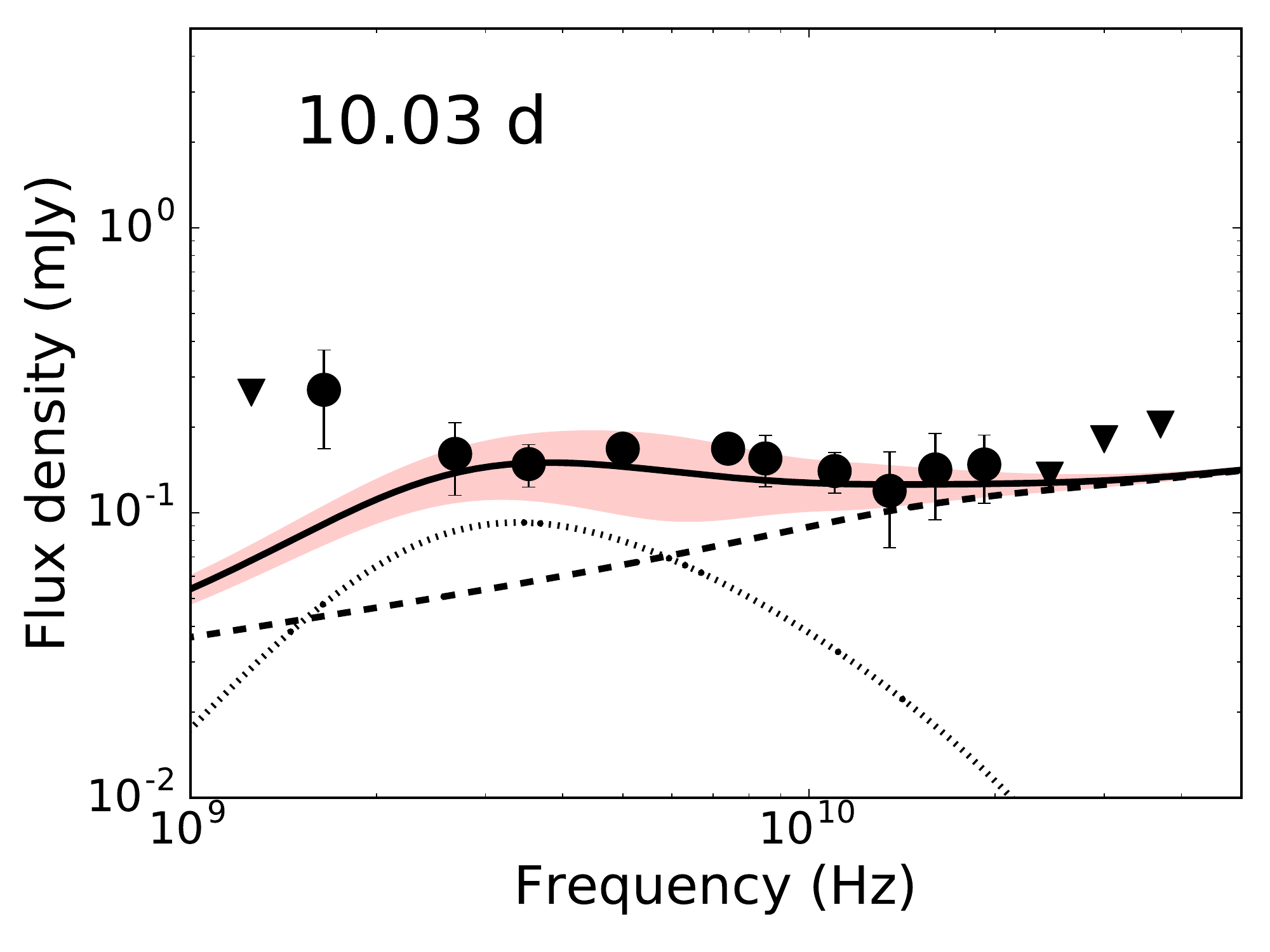} & 
 \includegraphics[width=0.31\textwidth]{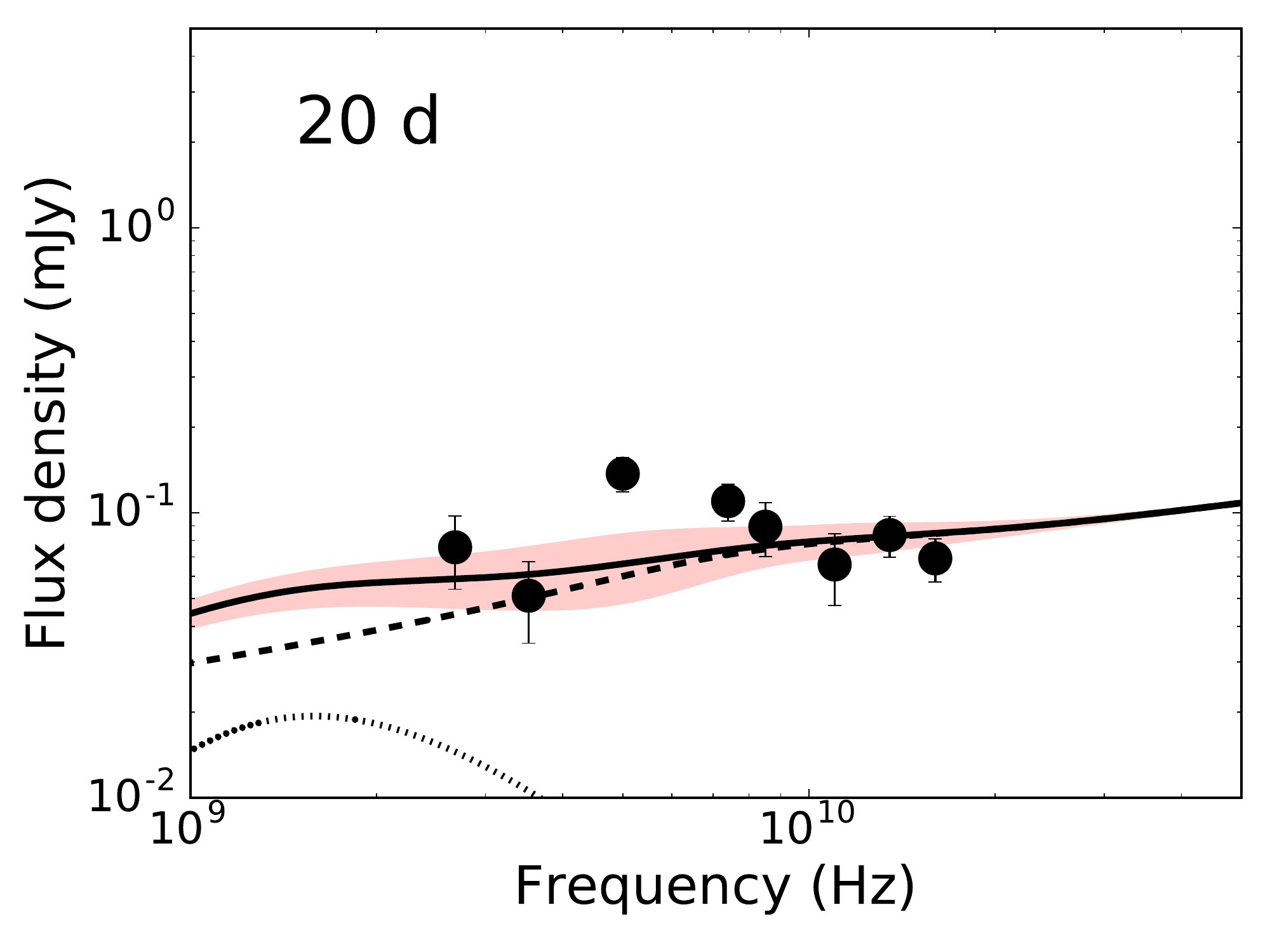} \\
\end{tabular}
\caption{Radio spectral energy distributions of the afterglow of \me\ at multiple epochs starting 
at 0.36~d, together with the same reverse shock (dotted) and forward shock (dashed) 
ISM model in Figure \ref{fig:160509A_radioxrtsed}. The red shaded regions represent the expected 
variability due to scintillation, which is greatest in the vicinity of the transition frequency 
along the line of sight to the GRB, $\nu_{\rm T} = 13.55$~GHz. The radio observations up to 10.03~d 
are dominated by the reverse shock.}
\label{fig:160509A_radioseds}
\end{figure*}

\begin{deluxetable}{lr}
\tabletypesize{\scriptsize}
\tablecaption{Model Parameters \label{tab:params}}
\tablehead{
\colhead{Parameter} & \colhead{Value}
}
\startdata
\multicolumn{2}{c}{Reverse Shock}  \\[2pt]
$\nuar$    & $2.5\times10^{10}$~Hz \\[2pt]
$\numr$    & $1.5\times10^{10}$~Hz \\[2pt]
$\nucr$    & $4\times10^{11}$~Hz   \\[2pt]
$\fnumaxr$ & $9$~mJy               \\[2pt]
\hline\\[-4pt]
\multicolumn{2}{c}{Forward Shock (ISM)}\\[2pt]
$p$ & $2.39\pm0.03$\\[2pt]
$\epse$ & $0.84^{+0.06}_{-0.08}$\\[2pt]
$\epsb$ & $0.11^{+0.07}_{-0.05}$\\[2pt] 
$\dens$ & $(8.6\pm2.2)\times10^{-4}$~\pcc\\[2pt]
$\EKiso$ & $\left(18.7^{+5.4}_{-2.6}\right)\times10^{52}$~erg\\[2pt]
$\AV$ & $3.35^{+0.08}_{-0.07}$~mag\\[2pt]
$\tjet$ & $5.7^{+0.6}_{-0.5}$~d \\[2pt]
$f_{\nu, \rm host, \it g}$ & $0.29~\mu$Jy\\[2pt]
$f_{\nu, \rm host, \it r}$ & $0.88~\mu$Jy\\[2pt]
$f_{\nu, \rm host, \it z}$ & $9.0~\mu$Jy \\[2pt]
$f_{\nu, \rm host, \it J}$ & $11.9~\mu$Jy \\[2pt]
$f_{\nu, \rm host, \it K}$ & $28.8~\mu$Jy \\[2pt]
$\thetajet$ & $3.89^{+0.14}_{-0.16}\degr$\\[2pt]
$\EK$\tablenotemark{a} & $\left(4.4^{+1.1}_{-0.7}\right)\times10^{50}$~erg\\[2pt]
$\Egamma$\tablenotemark{a,b} & $(1.3\pm0.1)\times10^{51}$~erg\\[2pt] 
$\nuaf$    & $1.2\times10^{7}$~Hz  \\[2pt]
$\numf$    & $8.7\times10^{14}$~Hz \\[2pt]
$\nucf$    & $3.2\times10^{15}$~Hz \\[2pt]
$\fnumaxf$ & $1.6$~mJy\\[2pt]
\hline\\[-4pt]
\multicolumn{2}{c}{Forward Shock (wind)}\\
$p$      & $2.11$ \\[2pt]
$\epse$  & $0.60$\\[2pt]
$\epsb$  & $0.40$\\[2pt]
$\Astar$ & $5.3\times10^{-3}$~\pcc\\[2pt]
$\EKiso$ & $3.0\times10^{53}$~erg\\[2pt]
$\AV$    & $4.1$~mag\\[2pt]
$\tjet$  & $5.5$~d\\[2pt]
$f_{\nu, \rm host, \it g}$ & $0.26~\mu$Jy\\[2pt]
$f_{\nu, \rm host, \it r}$ & $0.86~\mu$Jy\\[2pt]
$f_{\nu, \rm host, \it z}$ & $7.2~\mu$Jy \\[2pt]
$f_{\nu, \rm host, \it J}$ & $15.7~\mu$Jy \\[2pt]
$f_{\nu, \rm host, \it K}$ & $66.4~\mu$Jy \\[2pt]
$\thetajet$ & $1.6\degr$\\[2pt]
$\EK$\tablenotemark{a} & $1.3\times10^{50}$~erg\\[2pt]
$\Egamma$\tablenotemark{a,b} & $(2.2\pm0.2)\times10^{50}$~erg\\[2pt] 
$\nuaf$    & $1.2\times10^7$~Hz\\[2pt]
$\numf$    & $1.2\times10^{14}$~Hz\\[2pt]
$\nucf$    & $1.1\times10^{16}$~Hz\\[2pt]
$\fnumaxf$ & $1.6$
\enddata
\tablecomments{All frequencies and flux densities in this table are calculated at 1~d. The host 
flux density measurements are corrected for Milky Way extinction and are presented for a 
representative model.}
\tablenotetext{a}{Corrected for beaming.}
\tablenotetext{b}{1--$10^4$~keV, rest frame.}
\end{deluxetable}

\subsection{The Reverse Shock}
\label{text:RS}
We construct a model SED for the radio to X-ray emission at 1.13~days comprising two emission 
components: (1) a FS (Section \ref{text:FS}), which peaks between the radio and optical bands, fits 
the NIR to X-ray SED, and provides negligible contribution in the radio band, and (2) a RS (this 
section), which fits the radio SED and provides negligible contribution at higher frequencies. The 
synchrotron parameters of the RS are 
listed in Table \ref{tab:params}.
We find that this combined RS plus FS model 
completely describes the observed SED at 1.13 days (Figure 
\ref{fig:160509A_radioxrtsed}).

We evolve both emission components to the epochs of our radio observations. The evolution 
of the RS spectrum depends on whether the shock is Newtonian or relativistic in the frame of the 
unshocked ejecta, and is determined by the evolution of the ejecta Lorentz factor with radius, 
quantified by the parameter $g$: $\Gamma\propto R^{-g}\propto t^{-g/(1+2g)}$. This was first 
measured observationally for GRB~130427A, where a value of $g\approx5$ was inferred 
for a Newtonian RS \citep{lbz+13}. We find that evolving the RS SED for \me\ with $g \approx 2$ 
matches the observed radio spectrum well from 0.36~d to 10~d. This value of $g$ closely matches the 
predicted value of $g\approx2.2$ from numerical calculations of the RS evolution for a Newtonian RS 
\citep{ks00}. A value of $g\approx3$ expected for a relativistic RS is ruled out by the observed 
evolution of the radio SED, providing the second direct measurement of $g$, and the first 
observational confirmation of the numerical theory.

The radio peak ascribed to the RS emission fades faster than expected from the RS model after 
$\approx5$~d. We note that this coincides with the time of the jet break in the X-ray light curve 
(Section \ref{text:basic_considerations}). The standard FS jet break is a 
combination of geometrical effects that take place when the FS Lorentz factor, $\Gamma\approx 
\theta_{\rm jet}^{-1}$: the observer sees the edge of the jet and the swept-up material begins to 
expand sideways \citep{rho99,dcrgl12,gp12}. In the case of the RS, the ejecta internal energy drops 
rapidly after the RS crossing and the local sound velocity in the ejecta is expected to be 
sub-relativistic. Thus, we expect the lateral expansion to be fairly slow, resulting in no change in 
the dynamics or the scaling of the RS break frequencies across the jet break. The geometric effect 
is expected to dominate, resulting in a change in the RS peak flux scaling by $\Gamma_{\rm RS}^2$ at 
$\tjet$.
Setting the RS jet break time to 5.2~d as derived from a preliminary fit to the FS (Section 
\ref{text:FS}), we find that the resultant evolution of the RS SED fits all subsequent radio 
observations well (Figure \ref{fig:160509A_radioseds}).

Finally, we note that $\nucr$ passes through the NIR at $\approx 3\times10^{-2}$~d in this model. 
After this time, we do not expect observable RS emission in the optical/NIR. This is consistent with 
the earliest available $R$-band observation \cite[$R<19.5$~mag at 
$6.5\times10^{-2}$~d;][]{gcn19409}, and with all subsequent optical/NIR data. 

\subsection{The Forward Shock}
\label{text:FS}
To model the FS emission we employ the framework of synchrotron radiation from relativistic 
shocks, including the effects of inverse Compton cooling \citep{se01,gs02}. The parameters of 
the fit are the kinetic energy (\EKiso), the density (\dens), the electron energy index ($p$), and 
the fraction of the shock energy given to electrons (\epse) and magnetic fields (\epsb). We 
use the Small Magellanic Cloud (SMC) extinction curve to model the extinction (\AV) in the GRB host 
galaxy \citep{pei92}, and include the flux density of the host in the $grzJK$ bands ($f_{\nu, \rm 
host}$), together with the jet break time ($\tjet$), as additional free parameters. 

The afterglow observations in this case do not allow us to directly determine the circumburst 
density profile, and both ISM and wind-like environments have been inferred for GRBs in 
the past \citep[e.g.][]{pk02,yhsf03,cfh+10,cfh+11,skb+11}. However, we find that consistency 
arguments between the FS and RS SEDs at the deceleration time provide meaningful results in 
the ISM case, but not in the wind case. We therefore focus on the ISM model in the remainder of the 
article, and discuss the wind model briefly in Section \ref{text:wind}. 

\begin{figure*}
\begin{tabular}{ccc}
 \centering
 \includegraphics[width=0.31\textwidth]{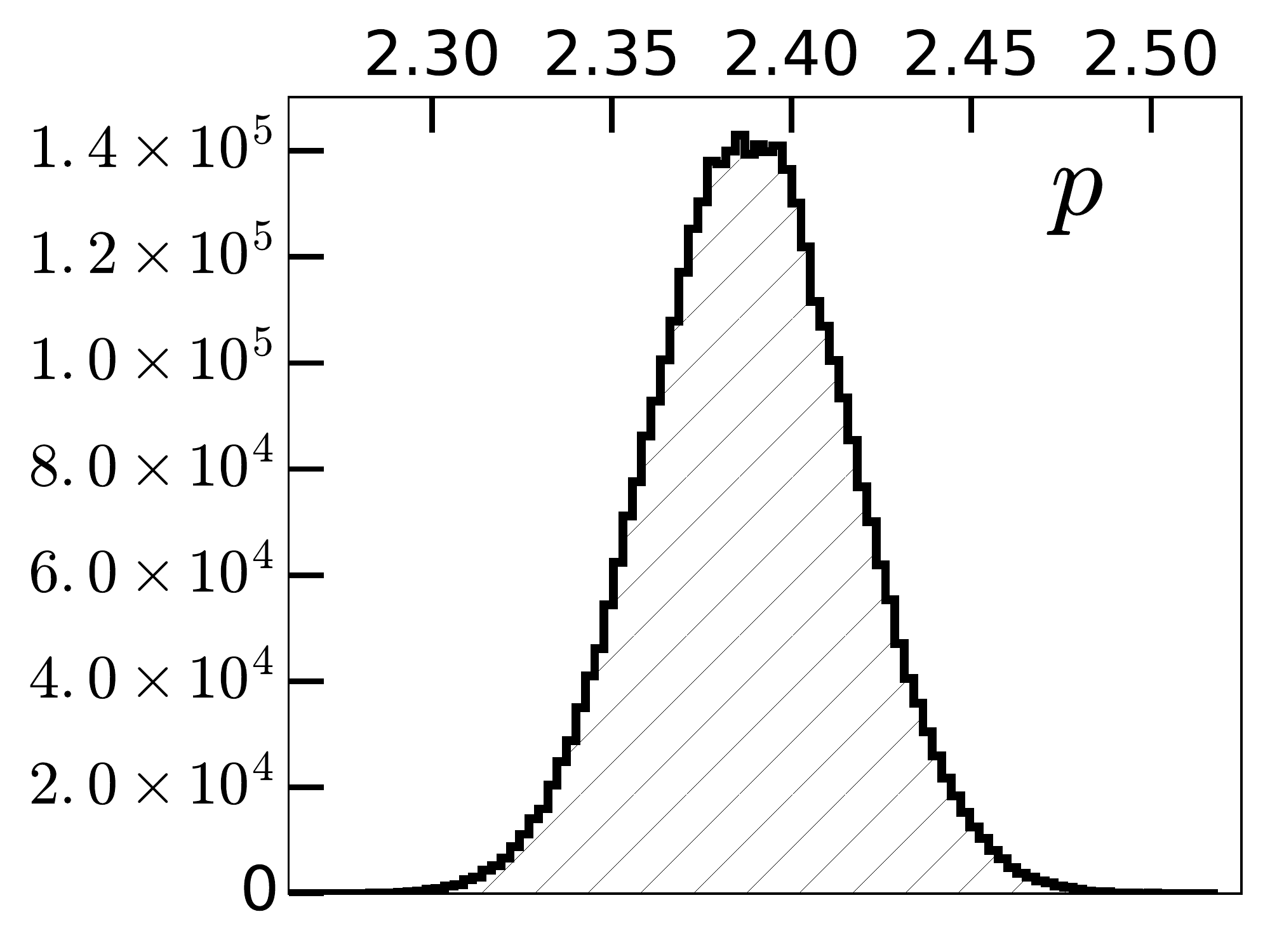} &
 \includegraphics[width=0.31\textwidth]{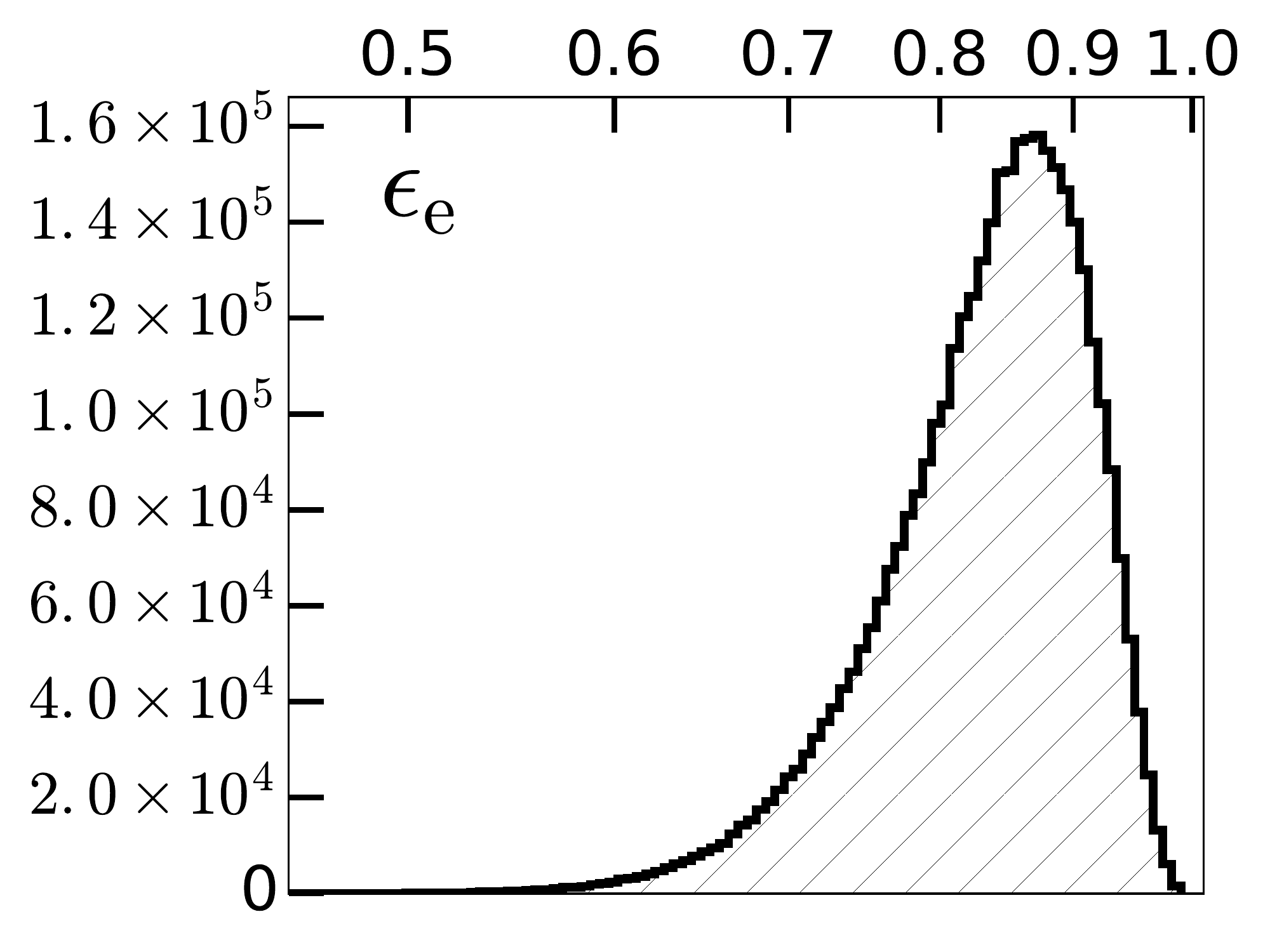} &
 \includegraphics[width=0.31\textwidth]{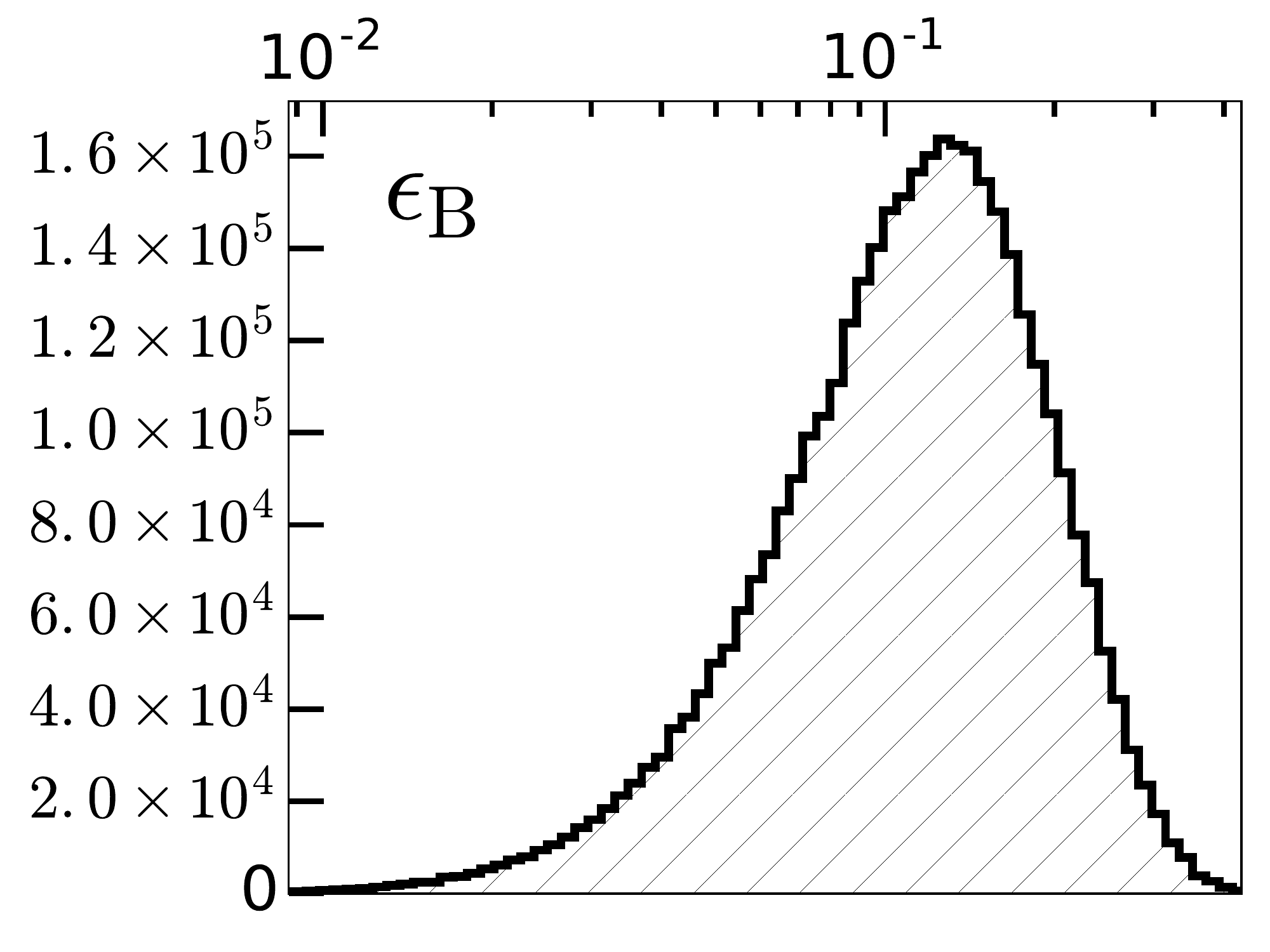} \\
 \includegraphics[width=0.31\textwidth]{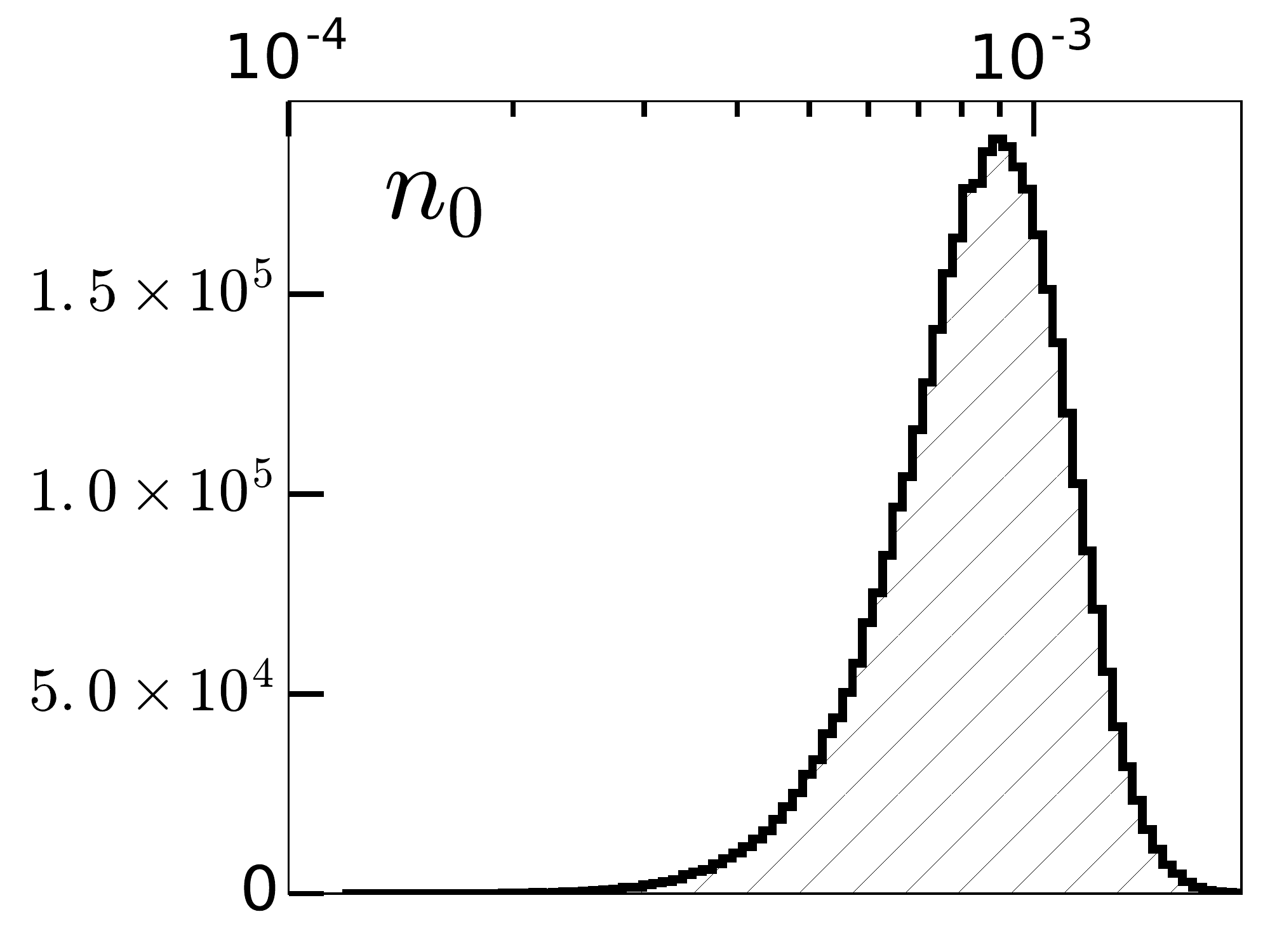} &
 \includegraphics[width=0.31\textwidth]{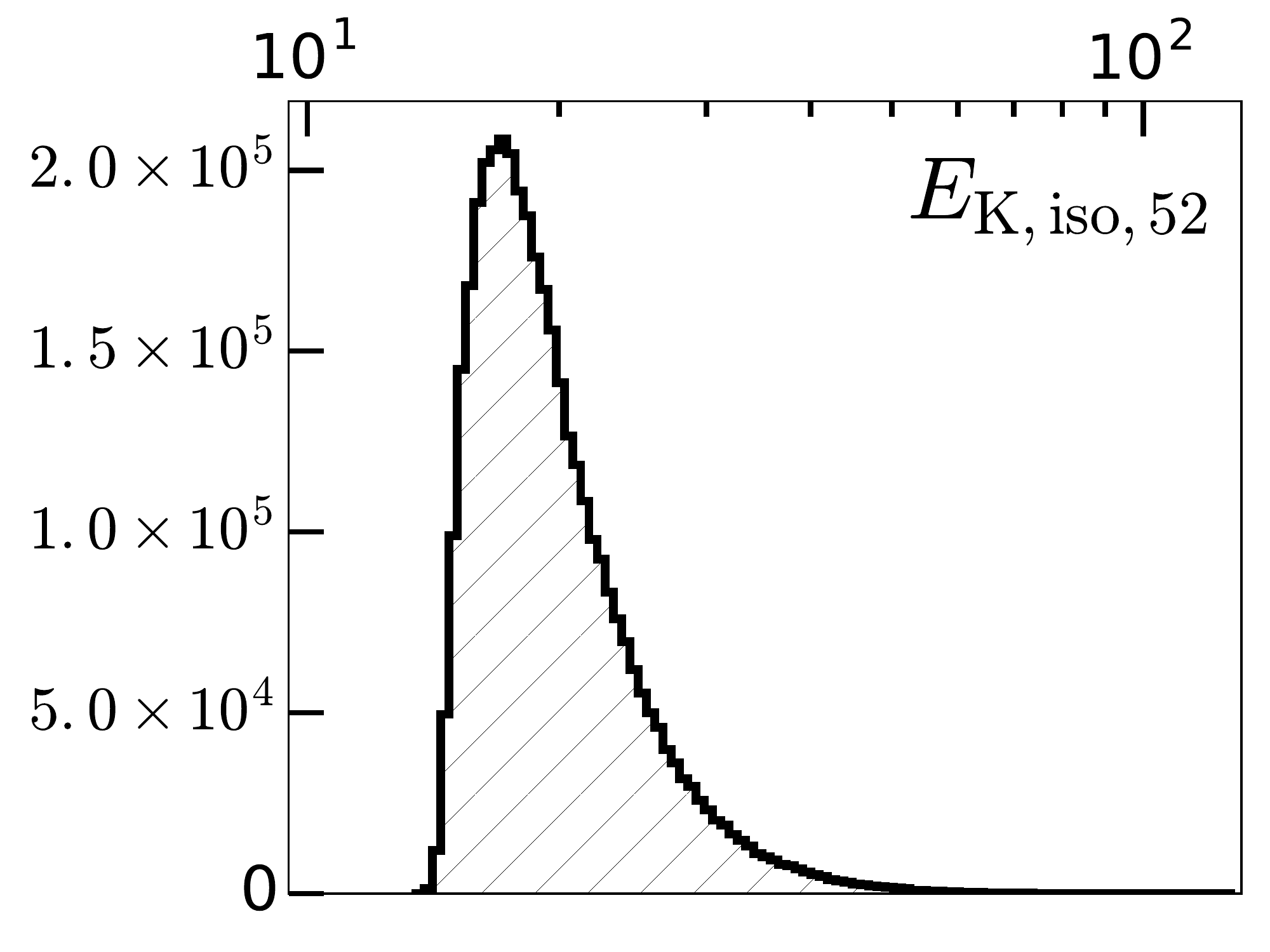} &
 \includegraphics[width=0.31\textwidth]{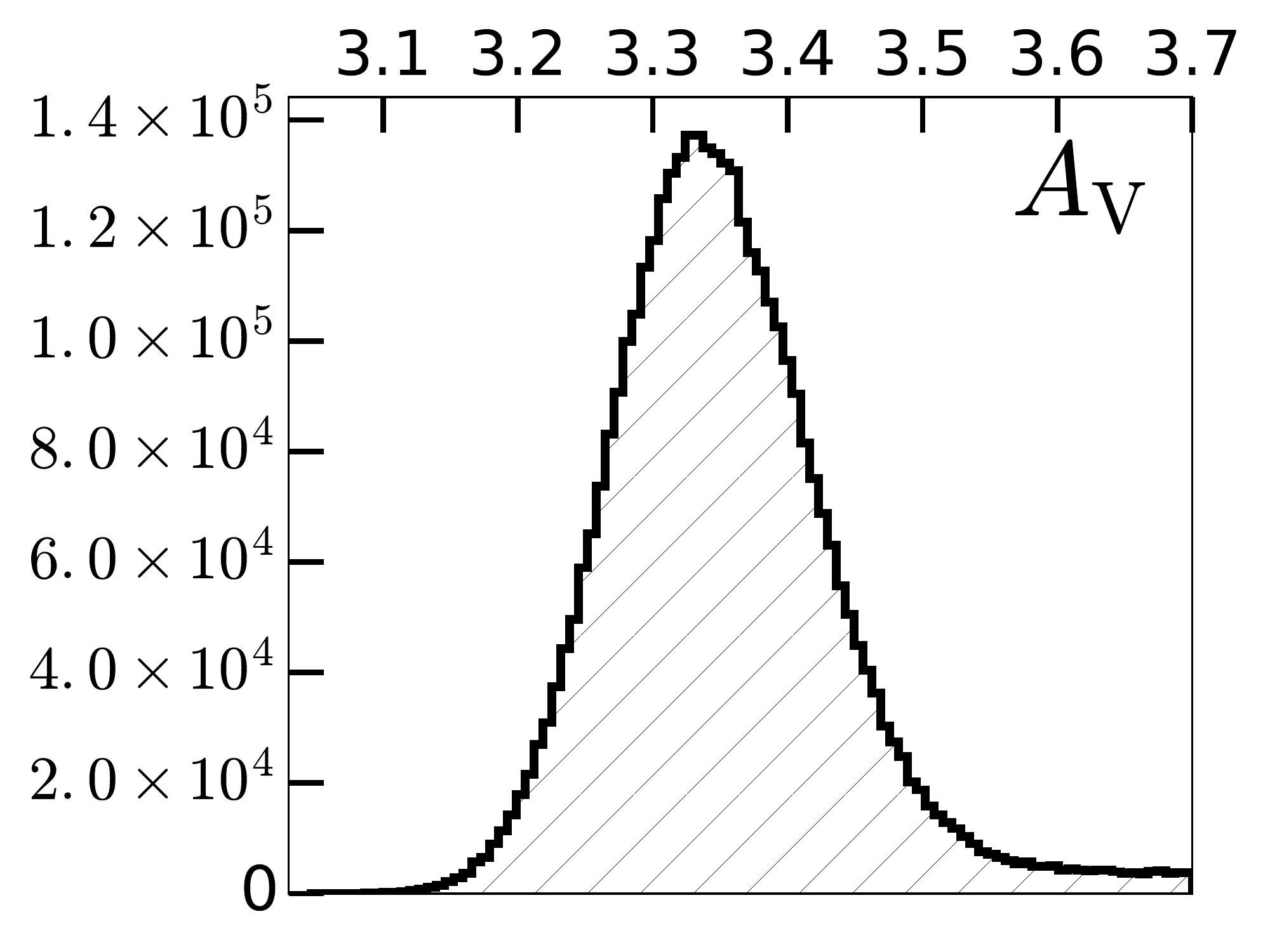} \\
 \includegraphics[width=0.31\textwidth]{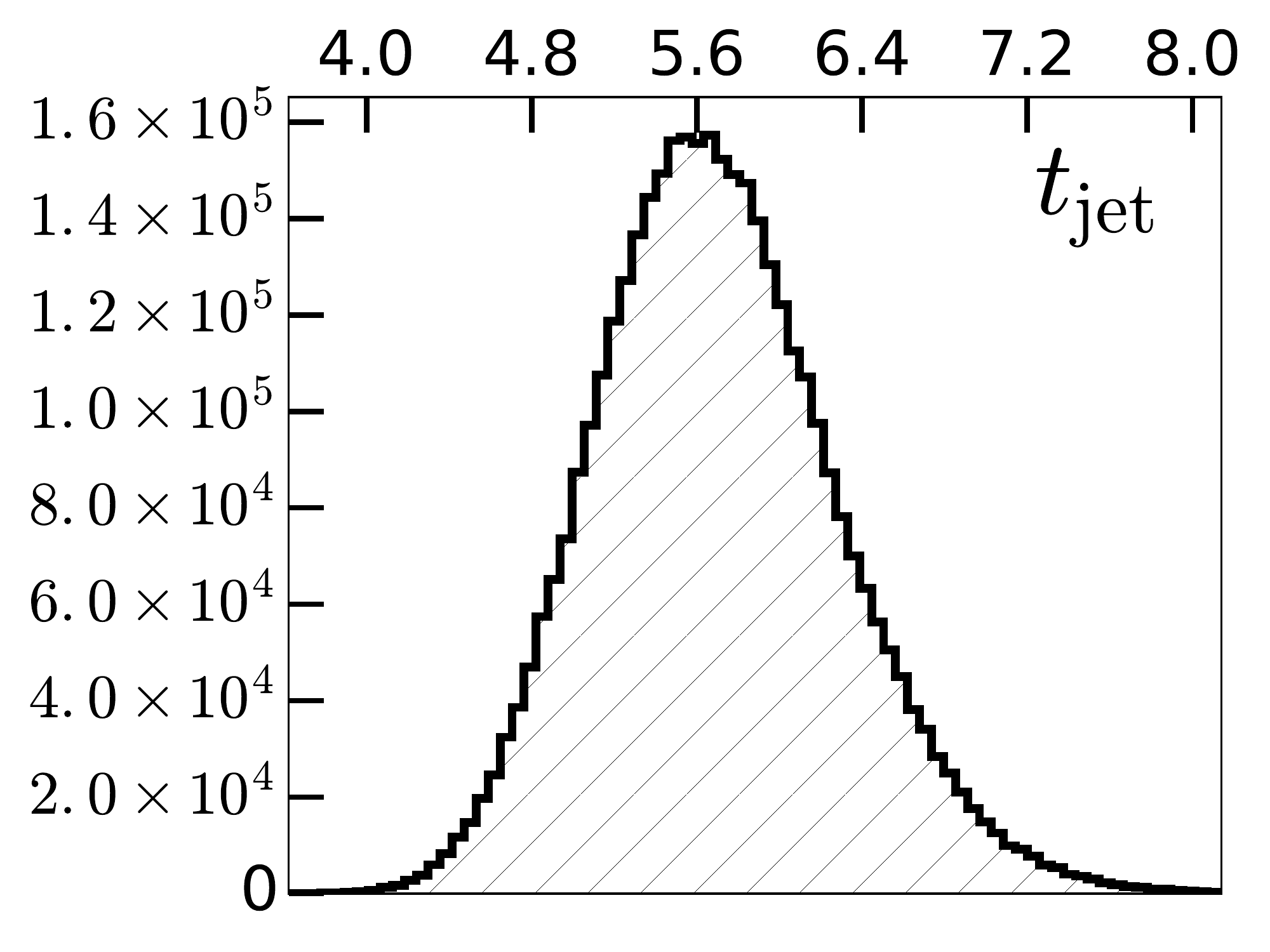} &
 \includegraphics[width=0.31\textwidth]{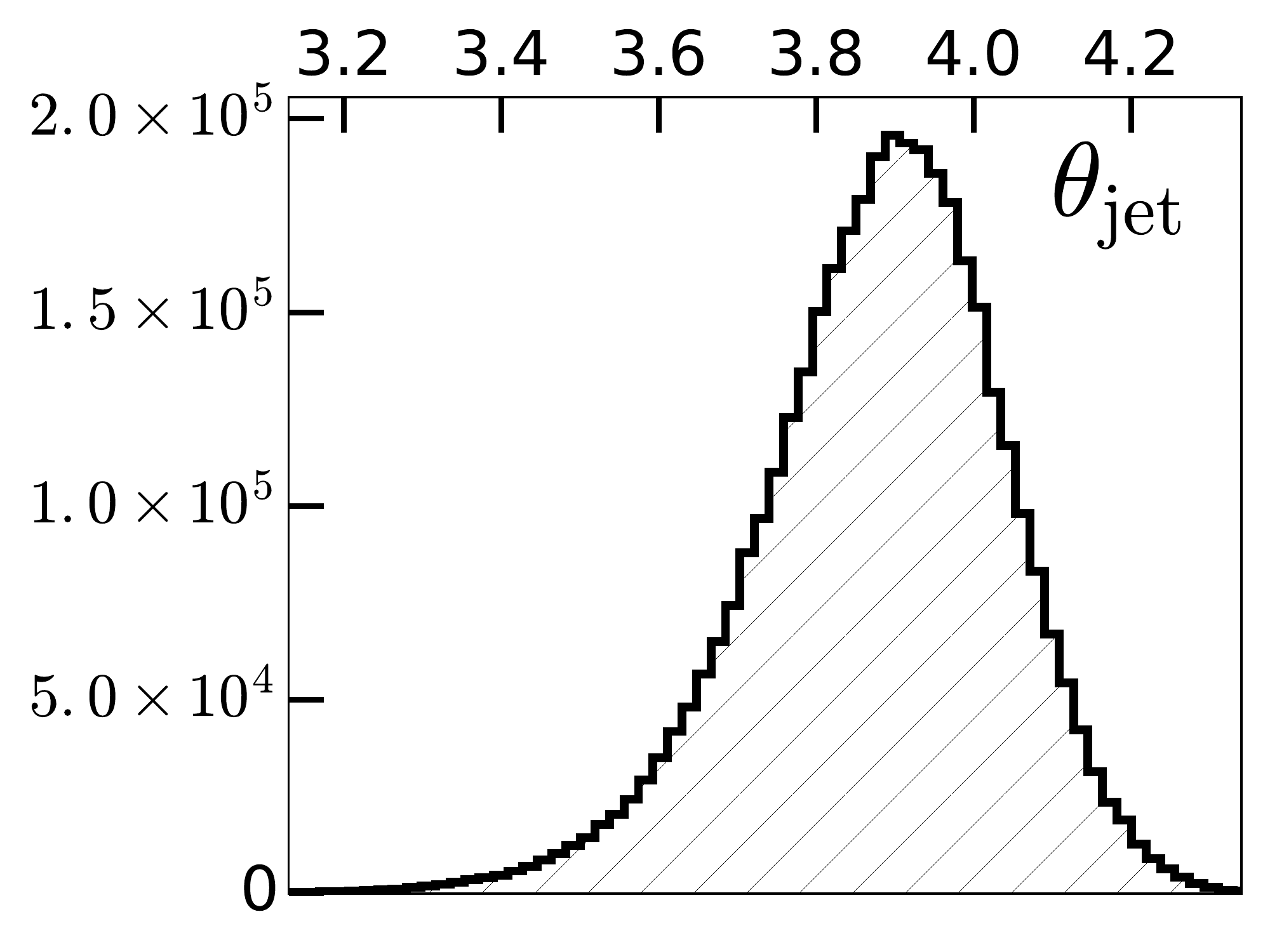} & 
 \includegraphics[width=0.31\textwidth]{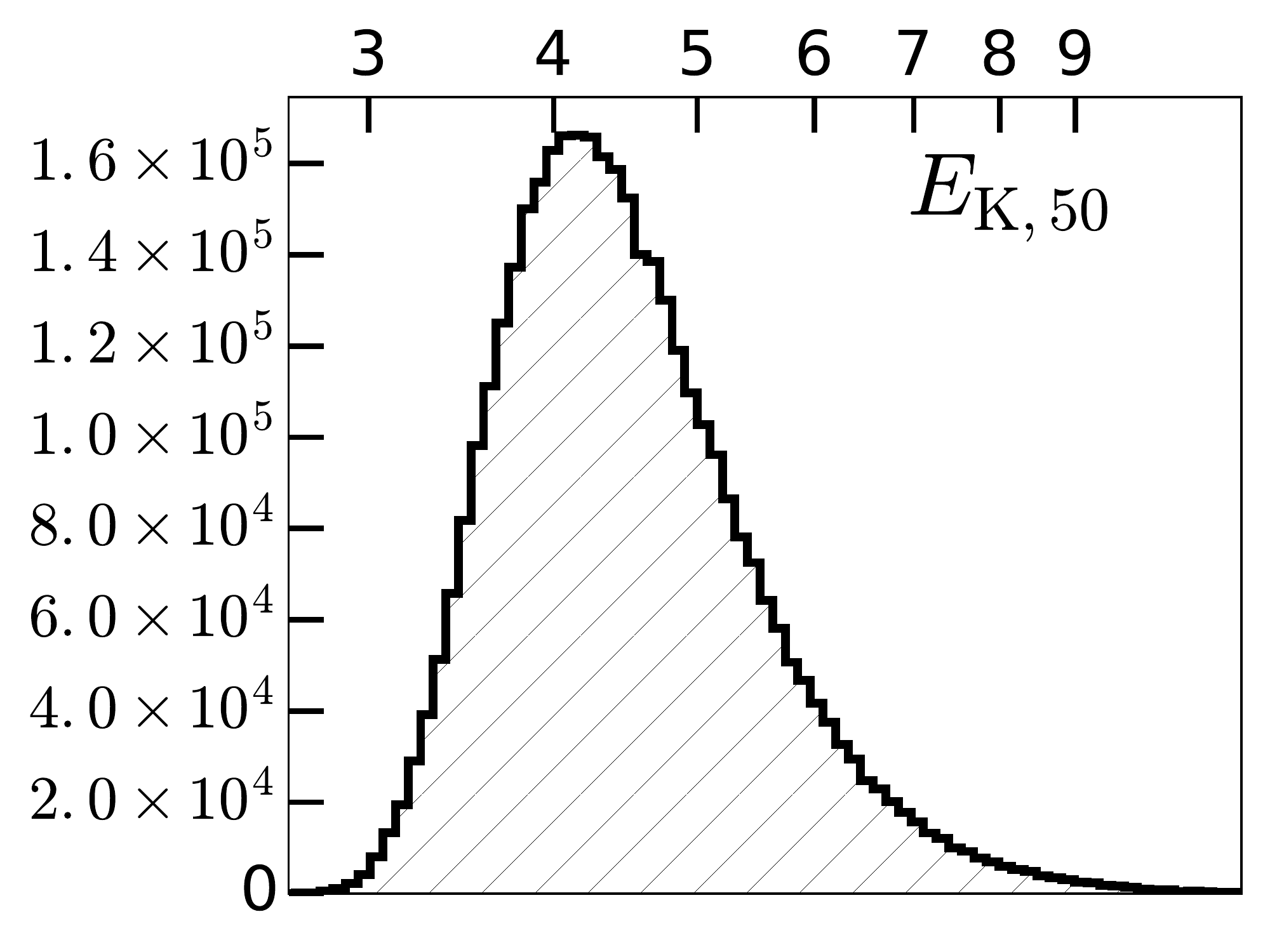}
\end{tabular}
\caption{Marginalized posterior probability density functions of the FS parameters from MCMC 
simulations. We have restricted $\epse+\epsb < 1$.}
\label{fig:hists}
\end{figure*}

\begin{figure*}
\begin{tabular}{ccc}
 \centering
 \includegraphics[width=0.31\textwidth]{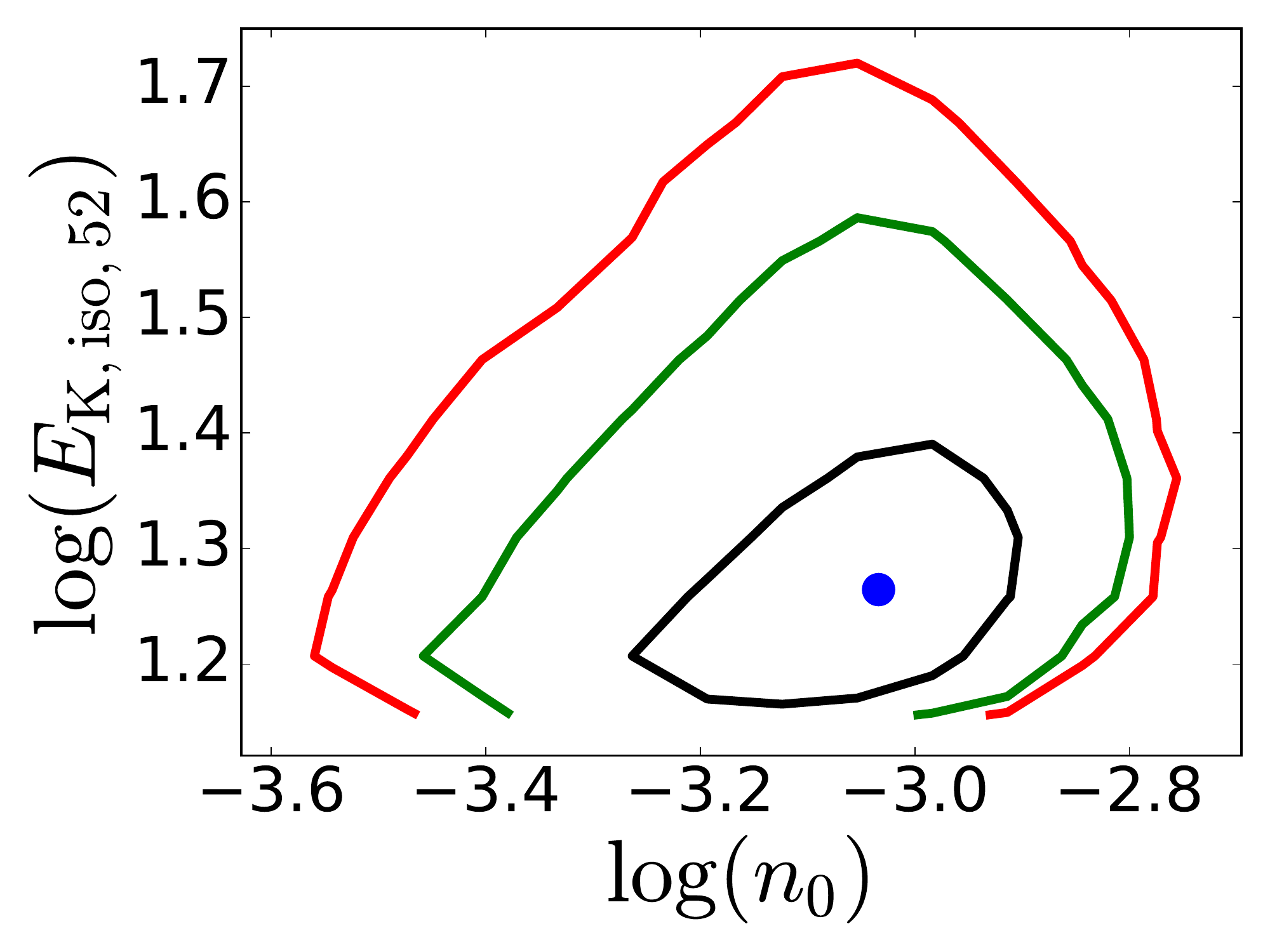} &
 \includegraphics[width=0.31\textwidth]{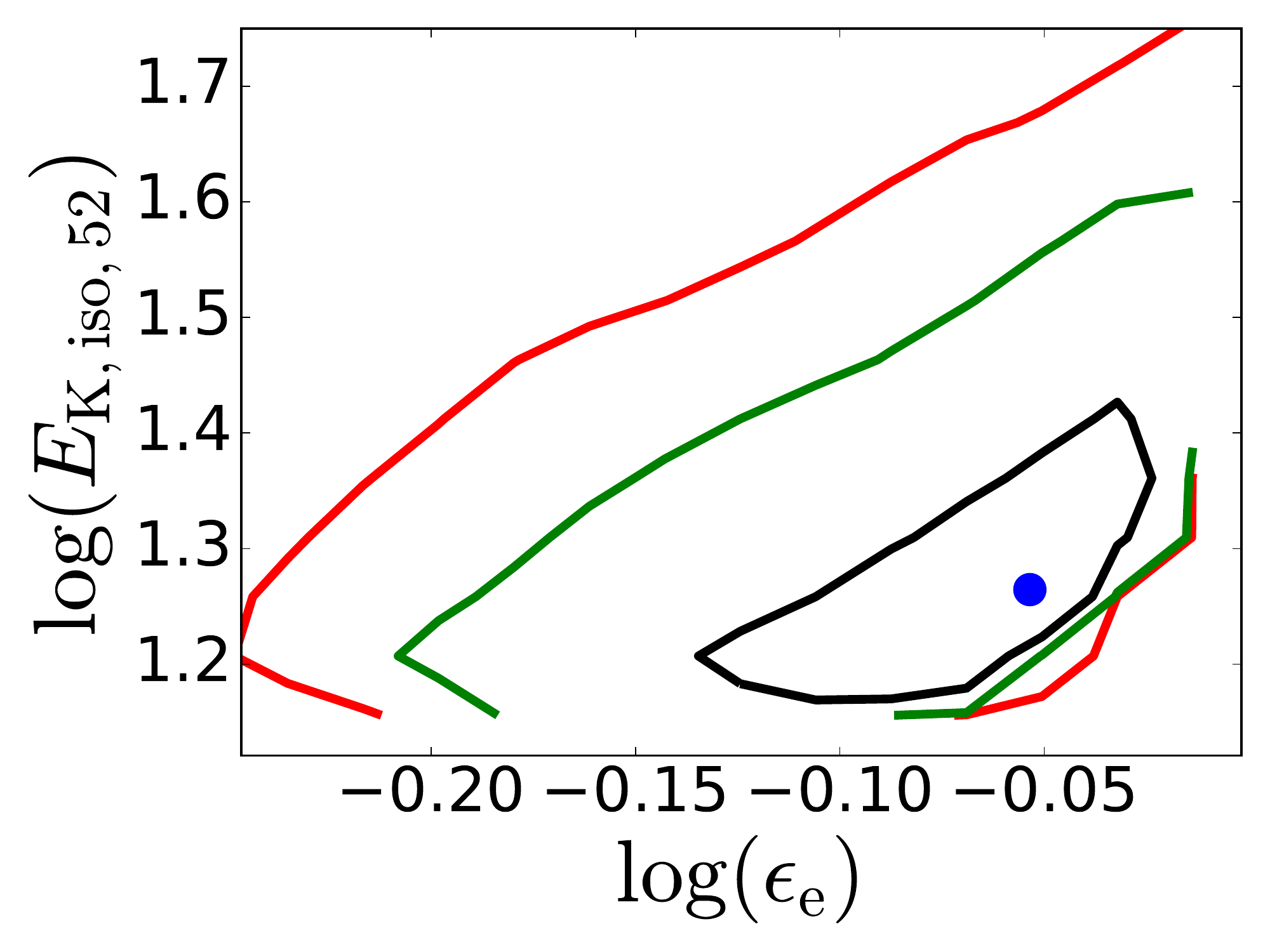} &
 \includegraphics[width=0.31\textwidth]{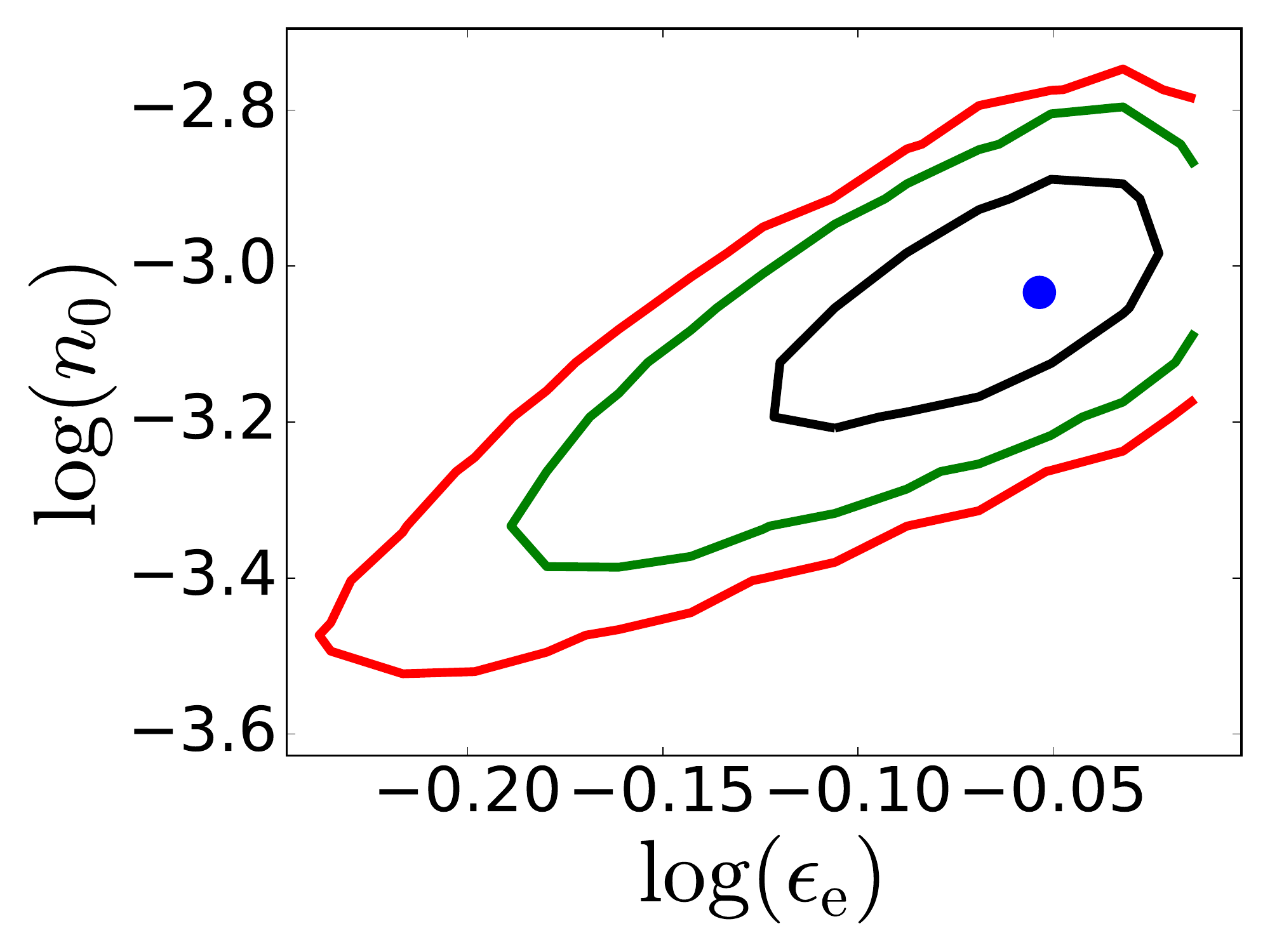} \\
 \includegraphics[width=0.31\textwidth]{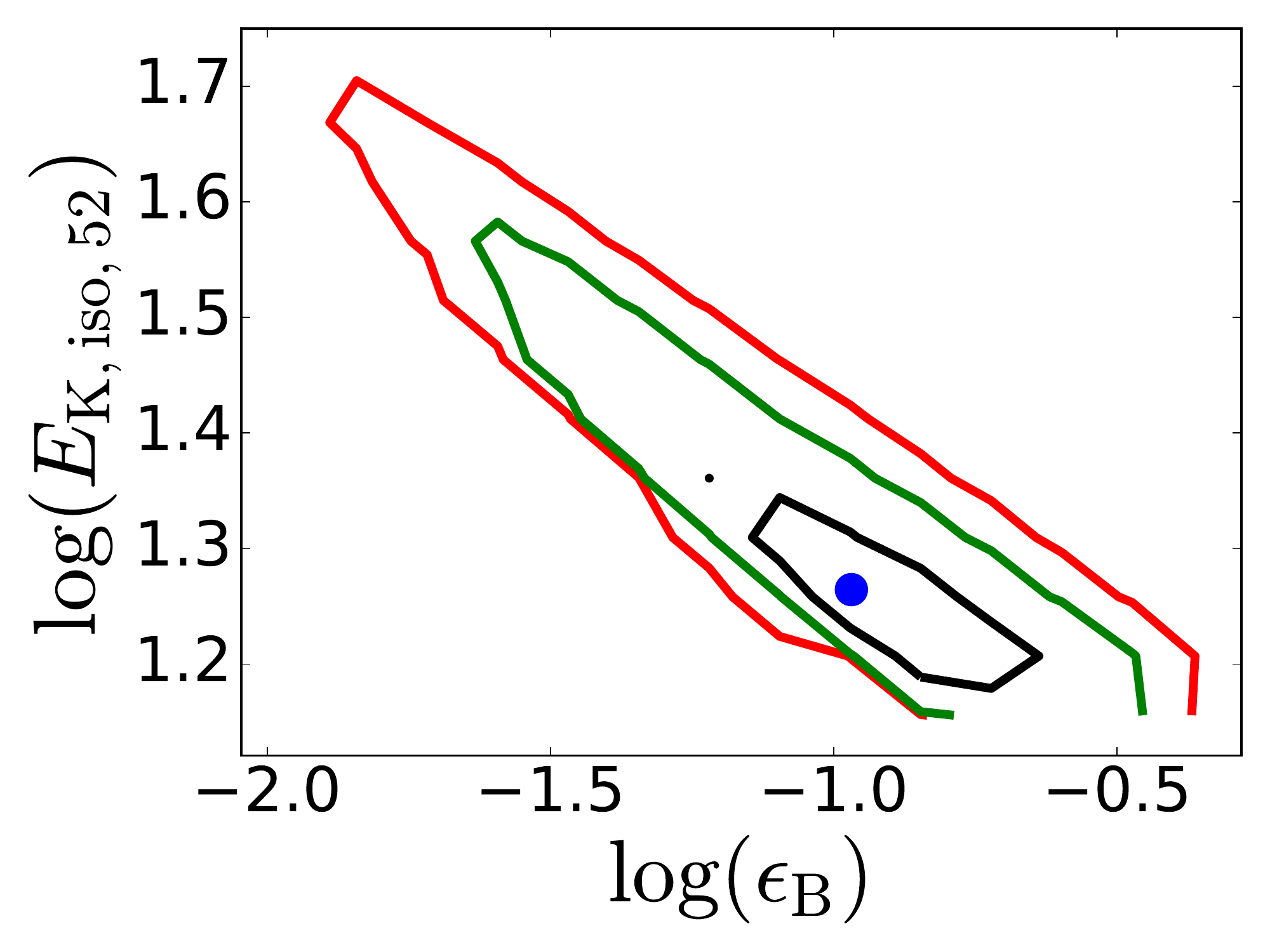} &
 \includegraphics[width=0.31\textwidth]{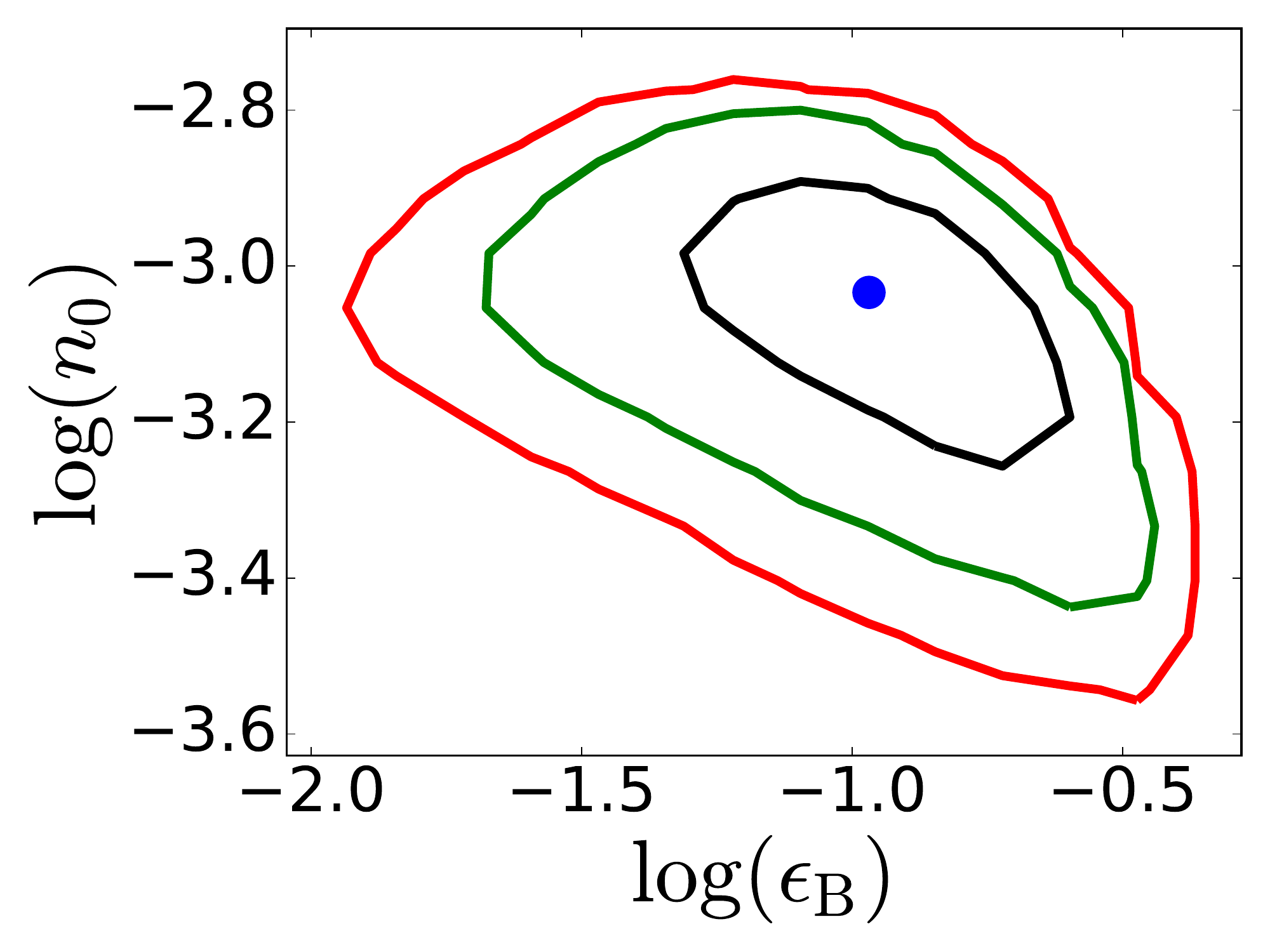} &
 \includegraphics[width=0.31\textwidth]{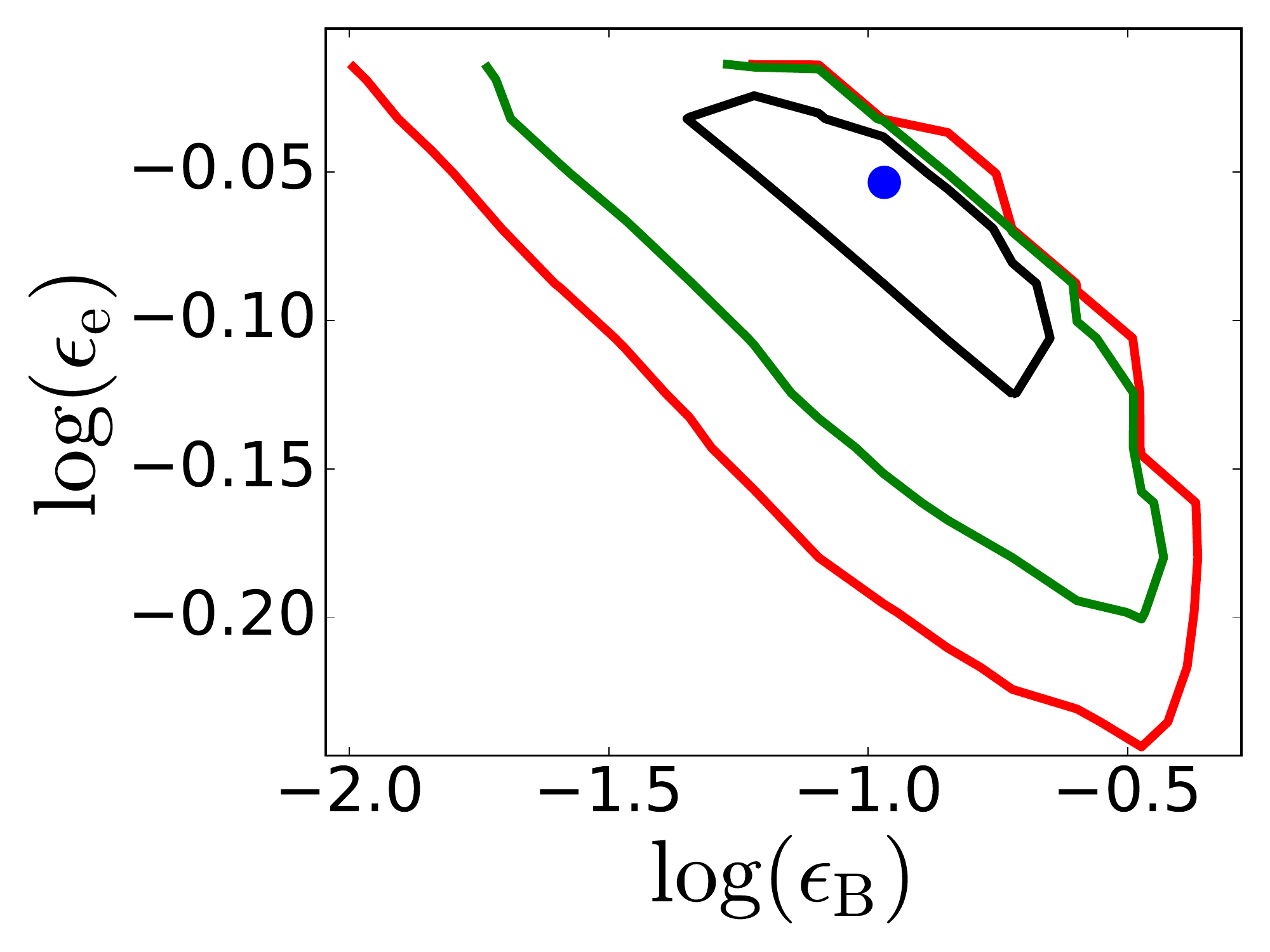} 
\end{tabular}
\caption{1$\sigma$ (red), 2$\sigma$ (green), and 3$\sigma$ (black) contours for correlations 
between the physical parameters \EKiso, \dens, \epse, and \epsb\ from Monte Carlo simulations,
together with the best-fit model (blue dot). 
We have restricted $\epse+\epsb<1$. 
}
\label{fig:corrplots}
\end{figure*}

We fit all available photometry with a combination of the RS and FS contributions. A least-squares 
analysis provides the starting point, using which we find a FS jet break time of 
$\tjet\approx5.2$~d.
We fix the RS jet break time to this value. To efficiently sample parameter space and to uncover 
correlations between the parameters, we then carry out a Markov Chain Monte Carlo (MCMC) analysis 
using \textsc{emcee} \citep{fhlg13}. Our analysis methods are described in detail by \cite{lbt+14}.
The resultant marginalized posterior density functions are summarized in Table \ref{tab:params} 
and Figure \ref{fig:hists}. Correlation functions between the four physical parameters 
are plotted in Figure \ref{fig:corrplots}. 
In our best-fit model ($\chi^2 = 16.4$ for 12 degrees of freedom), the FS transitions from fast 
cooling to slow cooling at $\approx0.3$~d, while
the Compton Y-parameter is $\approx2.4$, indicating that inverse-Compton cooling is moderately 
significant.

\section{Discussion}
\label{text:discussion}
\subsection{Self-consistency of RS and FS models}
\label{text:consistency}
In the standard synchrotron model, the break frequencies of the RS and FS spectra are expected to 
be related at \tdec: 
$\nucr/\nucf \sim \RB^{-3/2}$, $\numr/\numf \sim \RB^{1/2}\Gamma_0^{-2}$, and 
$\fnumaxr/\fnumaxf\sim\Gamma_0\RB^{1/2}$, where $\Gamma_0$ is the bulk Lorentz factor at \tdec, and 
$\RB\equiv\epsilon_{\rm B, RS}/\epsilon_{\rm B, FS}$ is the ejecta magnetization parameter 
\citep{gkg+08,hk13}. The three relations above then provide three constraints that can be solved 
exactly for \tdec, $\Gamma_0$, and \RB. For our best-fit FS+RS model, we find $\tdec \approx 
460$~s~$\approx T_{90}$, $\Gamma_0\approx 330$, and $\RB\approx8$. We note that the derived values 
of $\EKiso$, $\dens$, $\thetajet$, and $\Gamma_0$ can be used to derive a jet break time for the RS 
using the relation, $\tjet = 110(1+z)(E_{\rm 
K,iso,52}/\dens)^{1/3}\thetajet^{5/2}\Gamma_0^{-1/6}$~d \citep{glz+13}. Using the best-fit FS model, 
we find $t_{\rm jet, RS} \approx 3.4$~d, which is slightly earlier than the FS jet break time, as 
expected. The difference between this value and our assumed value of $\approx5.2$~d in Section 
\ref{text:RS} only marginally affects the fit at one of the epochs (4.06~d) in Figure 
\ref{fig:160509A_radioseds}. A fully consistent solution requires bootstrapping the FS and RS 
parameters together, and we defer such an analysis to future work.

\subsection{Low-density Environments and the RS}
In our previous work on GRB~130427A, we suggested that a slow-cooling RS is more likely to produce 
detectable radio emission \citep{lbz+13}. Since $\nucr/\nucf\propto\dens^{-4/3}$ at \tdec, a 
low-density environment may be a requisite factor for observing long-lasting RS emission 
\citep{kob00,rz16}. We find a low circumburst density in the context of long-lasting reverse 
shock emission for \me, leading credence to this  hypothesis. However, we also note that additional 
considerations such as high $\fnumaxr$ or late deceleration times may also contribute to stronger RS 
signatures; therefore, the detectability of a RS remains a complex question \citep{kmk+15}.


\subsection{Wind Model}
\label{text:wind}
Since the available afterglow observations do not distinguish strongly between a wind and ISM 
model, we also provide the parameters for a fiducial wind model (Table \ref{tab:params}).
For this model, the spectrum transitions from fast cooling to slow cooling at 0.17~d, and the 
spectral break frequencies at 1~d are 
within a factor of $\approx 3$ of the values derived for the ISM model in Section 
\ref{text:FS}. We note that the value of $g\approx2$ for the RS remains plausible in the 
wind environment as well and, therefore, the RS parameters derived in Section \ref{text:RS} 
remain reasonable. Combining the RS and FS parameters for the wind model, we find $\tdec 
\approx 170$~s, $\Gamma_0\approx34$, and $\RB\approx0.05$. The low value of $\Gamma_0$, the 
low inferred magnetization, and finding $\tdec\lesssim T_{90}$, together argue against the wind 
model \citep{feh93,wl95}. 

\subsection{Neutral Hydrogen Column Density and Extinction}
A correlation between the neutral hydrogen column derived from X-ray absorption and the 
line-of-sight extinction, $N_{\rm H} \approx 2\times10^{21}\pcmsq(\AV/{\rm mag})$, has been 
observed for the Milky Way \citep{ps95,go09}.
However, the majority of GRB afterglows exhibit lower values of $\AV$ than would be 
expected from this correlation \citep[e.g.,][]{gw01,sfa+04,zwm+10,zbm+13a}.
We note that the extinction of GRB afterglows by their host galaxy is often well fit with an SMC 
extinction curve \citep[as we also do here;][]{jcg+15}. We therefore derive a corresponding 
correlation for the SMC using the relation between $N_{\rm H}$ and $E(B-V)$ from \cite{wxw12} and 
the mean $R_{\rm V}=2.74$ for the SMC bar from \cite{gcm+03}, obtaining $\log{\left(N_{\rm 
H}/10^{21}\pcmsq\right)} = 21.95\pm0.36 + \log\left(\AV/{\rm mag}\right)$. 
For $N_{\rm H} \approx 1.5\times10^{22}$\pcmsq, this gives $\log(\AV/{\rm mag}) = 0.23\pm0.36$ 
or $\AV=1.7^{+2.2}_{-1.0}$~mag, while the MW correlation gives $A_{\rm V}=(7.6\pm0.7)$~mag. Our 
observed value of $\AV = 3.35^{+0.08}_{-0.07}$~mag is, therefore, intermediate between the values 
expected from the two relations.

\section{Conclusions}
\label{text:conclusions}
We present a detailed multi-wavelength study of the \Fermi-LAT GRB 160509A at $z = 1.17$. Our VLA 
observations spanning $0.36$--$20$ days after the burst clearly reveal the presence of multiple 
spectral components in the radio afterglow. We identify the two spectral components as arising from 
the forward and reverse shock, and from a joint analysis of the two emission components, we 
conclude:
\begin{itemize}[itemsep=5pt,topsep=5pt]
 \item The reverse shock dominates in the radio before $\approx 10$~d, and 
the forward shock dominates in the X-ray and optical/NIR.
\item The evolution of the reverse shock spectrum requires a Lorentz factor index, $g\approx2$, 
consistent with theoretical predictions for a Newtonian RS. We derive a deceleration time of 
$460$~s, a Lorentz factor of $\Gamma_0\approx330$ at the deceleration time, and an ejecta 
magnetization of $\RB\approx8$. 
\item The afterglow observations do not strongly constrain the density profile of the circumburst 
environment. However, the RS-FS consistency relations yield a very low Lorentz factor in the wind 
environment. 
\item We derive a circumburst density of $\dens\approx10^{-3}$\pcc, supporting the hypothesis that 
a low density environment may be a requisite factor in producing a slow-cooling and long-lasting RS.
\end{itemize}
This work follows on our previous successful identification and 
characterization of a reverse shock in GRB~130427A, and highlights the importance of 
rapid-response radio observations in the study of the properties and dynamics of GRB ejecta.

\acknowledgements 
T.L.~is a Jansky Fellow of the National Radio Astronomy Observatory. E.B.~acknowledges 
support from NSF grant AST-1411763 and NASA ADA grant NNX15AE50G. W.F.~is supported by NASA 
through Einstein Postdoctoral Fellowship grant number PF4-150121.
A.V.F.'s group at UC Berkeley has received generous financial assistance from Gary and Cynthia 
Bengier, the Richard and Rhoda Goldman Fund, the Christopher R. Redlich Fund, the TABASGO 
Foundation, NSF grant AST-1211916, and NASA/{\it Swift} grant NNX12AD73G. This work was supported in 
part by the NSF under grant No. PHYS-1066293; A.V.F. thanks the Aspen Center for Physics for its 
hospitality during the black holes workshop in June 2016. 
This research has made use of data obtained through the High Energy Astrophysics Science Archive 
Research Center Online Service, provided by the NASA/Goddard Space Flight Center. Some of the 
data presented herein were obtained at the W.M. Keck Observatory, which is operated as a scientific 
partnership among the California Institute of Technology, the University of California and the 
National Aeronautics and Space Administration, and was made possible by the generous financial 
support of the W.M. Keck Foundation. VLA observations were taken as part of our VLA Large Program 
15A-235 (PI: E.~Berger). The National Radio Astronomy Observatory is a facility of the National 
Science Foundation operated under cooperative agreement by Associated Universities, Inc.

\bibliographystyle{apj}


\end{document}